\documentclass[useAMS,usenatbib]{mn2e}
\usepackage{amsmath}
\usepackage{amssymb}
\usepackage{multirow}
\usepackage{graphicx}
\usepackage{epstopdf}

\def\simlt{\lower.5ex\hbox{$\; \buildrel < \over \sim \;$}}
\def\simgt{\lower.5ex\hbox{$\; \buildrel > \over \sim \;$}}

\newcommand{\bd}{\begin{displaymath}}
\newcommand{\ed}{\end{displaymath}}
\newcommand{\be}{\begin{equation}}
\newcommand{\ee}{\end{equation}}
\newcommand{\beqa}{\begin{eqnarray}}
\newcommand{\eeqa}{\end{eqnarray}}

\title[Constraining 21 cm with the X-ray background] {Constraining the
  redshifted 21-cm signal with the unresolved soft X-ray background}
\author[Fialkov, Cohen, Barkana \& Silk] {Anastasia
  Fialkov$^{1,2}$\thanks{E-mail: anastasia.fialkov@cfa.harvard.edu},
  Aviad Cohen$^{3}$,  Rennan Barkana$^{3,4,5,6}$ \& Joseph Silk$^{5,7,8}$  \\
  $^{1}$ Harvard-Smithsonian Center for Astrophysics, Institute for Theory and Computation, 60 Garden Street, Cambridge, MA 02138, USA\\
  $^{2}$ Departement de Physique, Ecole Normale Sup\'{e}rieure, CNRS,
  24 rue Lhomond, Paris, 75005 France\\
  $^{3}$ Raymond and Beverly Sackler School of Physics and Astronomy,
  Tel Aviv University, Tel Aviv 69978, Israel\\
  $^{4}$ Sorbonne Universit\'{e}s, Institut Lagrange de Paris (ILP),
  Institut d'Astrophysique de Paris, UPMC Univ Paris
  06/CNRS \\
  $^{5}$ Department of Astrophysics, University of Oxford, Denys
  Wilkinson
  Building, Keble Road, Oxford OX1 3RH, UK \\  
    $^{6}$ Perimeter Institute for Theoretical Physics, 31 Caroline St N., Waterloo, ON N2L 2Y5, Canada\\  
  $^{7}$ Department of Physics and Astronomy, The Johns Hopkins
  University, Baltimore, MD 21218, USA\\
   $^{8}$ Sorbonne Universit´es, UPMC Univ Paris 6 et CNRS, UMR 7095, Institut d’Astrophysique de Paris, 98 bis bd Arago, 75014 Paris, France}

\begin{document}
\pagerange{\pageref{firstpage}--\pageref{lastpage}} \pubyear{2015}
\maketitle

\label{firstpage}

\begin{abstract}
 We use the observed unresolved cosmic X-ray background (CXRB) in
  the $0.5-2$ keV band and existing upper limits on the 21-cm power
  spectrum to constrain the high-redshift population of X-ray sources,
  focusing on their effect on the thermal history of the Universe and
  the cosmic 21-cm signal. Because the properties of these sources are
  poorly constrained, we consider hot gas, X-ray binaries and
  mini-quasars (i.e., sources with soft or hard X-ray spectra) as
  possible candidates. We find that (1) the soft-band CXRB sets an
  upper limit on the X-ray efficiency of sources that existed before
  the end of reionization, which is one-to-two orders of magnitude
  higher than typically assumed efficiencies, (2) hard sources are
  more effective in generating the CXRB than the soft ones, (3) the
  commonly-assumed limit of saturated heating is not valid during the
  first half of reionization in the case of hard sources, with any
  allowed value of X-ray efficiency, (4) the maximal allowed X-ray
  efficiency sets a lower limit on the depth of the absorption trough
  in the global 21-cm signal and an upper limit on the height of the
  emission peak, while in the 21-cm power spectrum it sets a minimum
  amplitude and frequency for the high-redshift peaks, and (5) the
  existing upper limit on the 21-cm power spectrum sets a lower limit
  on the X-ray efficiency for each model. When combined with the 21-cm
  global signal, the CXRB will be useful for breaking degeneracies and
  helping constrain the nature of high-redshift heating sources.
\end{abstract}


\section{Introduction}
\label{Sec:Intro}

The cosmic X-ray background (CXRB) was the first cosmic background
radiation to be discovered \citep{Giacconi:1962}; however, its origin
is not fully understood yet.  Deep extragalactic X-ray surveys have
resolved $\approx76-88 \%$ of the CXRB into point sources, with the
resolved fraction increasing at higher energies
\citep{Lehmer:2012}. It is now clear that the resolved CXRB is mostly
produced by the integrated emission from Active Galactic Nuclei (AGN)
at $z\leq 8$ and galaxies located at redshifts out to $z = 2.6$
\citep{Lehmer:2012}. However, a part of the CXRB remains unresolved
and its origin is still undetermined. The largest fraction of the
unresolved CXRB is in the soft ($0.5-2$ keV) band, which has been
explored by a number of X-ray missions \citep{McCammon:2002,
  Hickox:2006, Lehmer:2012}; the unresolved fraction is 24.3\% of the
total $8.15 \pm 0.58 \times 10^{-12}~\rm{erg
  ~cm}^{-2}\rm{s}^{-1}\rm{deg}^{-2}$ background measured by {\it
  Chandra} \citep{Lehmer:2012} in this band. Future missions such as
the proposed X-ray Surveyor \citep{Weisskopf:2015} and
Athena\footnote{http://www.the-athena-x-ray-observatory.eu/} should be
able to resolve this fraction, thus further constraining the
contribution of distant and faint X-ray sources.

At present, the nature of the unresolved fraction of the CXRB is still
debated. On the one hand, measurements of the X-ray background hint
that in the soft band normal galaxies may play an increasingly
important role \citep{Xue:2011}, likely dominating over AGN at fluxes
below the detection limit of {\it Chandra} ($\sim 5 \times
10^{-18}~\rm{erg ~cm}^{-2}\rm{s}^{-1}$ in this band). Extrapolating
the available data, \citet{Bauer:2004} predicted that the number
density of star-forming galaxies should overtake that of AGN at fluxes
just below $\sim 1\times 10^{-17}~\rm{erg ~cm}^{-2}\rm{s}^{-1}$. In
agreement, \citet{Lehmer:2012} found that the normal-galaxy number
counts rise rapidly at the faint end, compared to those of AGN, and
contribute $\approx 46\%$ of the number counts in the soft band found
in the {\it Chandra} Deep Field-South.  On the other hand, an
increasing number of faint and high-redshift quasars is being
discovered with infrared, optical and X-ray surveys \citep{Moran:2014,
  Marleau:2014, Lemons:2015}, supporting the idea that black holes may
have a stronger than expected contribution to the CXRB
\citep{Dijkstra:2004, Madau:2015}.

 Our understanding of the Universe, based on theoretical models
  and numerical simulations, suggests scenarios from which the excess
  of the soft X-ray background might naturally emerge. Current models
  show that the unresolved CXRB in the $0.5-2$ keV band can be
  produced by faint point sources, either AGN or galaxies, with a
  possible additional contribution produced by truly diffuse
  components such as the intergalactic medium (IGM). In particular,
  \citet{Dijkstra:2012} assumed a power-law X-ray spectral energy
  distribution (SED) of sources and varied their X-ray luminosity per
  unit star formation rate with redshift, showing that faint galaxies
  at redshifts up to $z=10$ can fully account for the unresolved
  portion of the CXRB; \citet{Cappelluti:2012} investigated the power
  spectrum of the unresolved CXRB in the soft band and interpreted the
  signal as a mixture of contributions from low-luminosity AGN ($\sim
  20\%$), galaxies up to $z = 10$ ($\sim 25\%$), and thermal emission
  of the IGM ($\sim 55\%$), where the contribution of galaxies was
  modeled with a power-law SED and normalized using the observed X-ray
  luminosity function; \citet{Madau:2015} showed that a population of
  high-redshift quasars and AGN with a piece-wise power-law SED
  \citep{Haardt:2012} can explain $\sim 60\% $ of the unresolved CXRB
  at $\sim 2$ keV. The latter possibility is particularly interesting
  since, according to \citet{Madau:2015}, such a population of sources
  could also fully reionize hydrogen and helium if their UV emissivity
  is normalized to the recent results of \citet{Giallongo:2015}.

Naturally, it is tempting to explain the excess in the soft CXRB as a
contribution of high-redshift X-ray sources \citep{Christian:2013,
  McQuinn:2012} in galaxies existing before the end of the epoch of
reionization (EoR) at $z_{\rm re}\sim 6-9$ \citep{Becker:2015,
  Planck:2015}.  These galaxies emit UV and X-ray photons which ionize
and heat the intergalactic gas.  While UV radiation has a small mean
free path in the neutral medium and, thus, reionizes gas close to the
source \citep{Wyithe:2003b}, X-ray photons travel hundreds of comoving
megaparsecs away, heating and partially ionizing the neutral gas far
from the sources \citep{Mirabel:2011}. In most of the currently
discussed models of reionization, UV photons are more efficient than
X-rays in ionizing the medium, and reionization proceeds inside-out
down to the scale of the H~II bubbles \citep{Barkana:2004}, while
X-rays only preheat the IGM. This picture, however, depends on the
relative normalization of the two processes and there are scenarios in
which X-rays are more efficient than the UV photons in driving
reionization, and it proceeds more homogeneously, with significant
smoothing up to the typical X-ray mean free path \citep{McQuinn:2012,
  Mesinger:2013, Majumdar:2015}.  The hardest X-rays emitted by a
population of high-redshift sources are expected to have such large
mean free paths that they are never absorbed by the IGM and, thus,
contribute to the unresolved soft CXRB observed today.

The extent to which the first X-ray sources contribute to the CXRB, as
well as their role in the thermal history and reionization of the IGM,
depends on their nature. At present these sources are poorly
  constrained due to the lack of observations and there are several
  possible candidates, including X-ray binaries (XRB)
  \citep{Power:2009, Mirabel:2011, Power:2013, Fragos:2013},
  mini-quasars \citep{Madau:2004}, and hot gas in galaxies,  hard
  photons emitted as a result of high-redshift supernovae activity
  \citep{Oh:2001}, as well as more exotic candidates such as
annihilating dark matter \citep{Cirelli:2009}. The spectral energy
distribution of X-ray photons emitted in each case depends on the
character of the sources, varying from a hard spectrum that peaks at
the photon energy of $\sim 3$ keV, as in the currently favored case of
XRBs \citep{Mirabel:2011, Fragos:2013, Fialkov:2014}, to a soft
power-law SED expected from hot gas heated by supernova explosions and
winds within galaxies (e.g., \citet{Furlanetto:2006b}).

The details of cosmic heating and reionization affect the
  redshifted 21-cm signal of neutral hydrogen (e.g., see
  \citet{Furlanetto:2006, Pritchard:2007, Ripamonti:2008,
    Pritchard:2012}). This radio signal, which is hoped to be the
richest future probe of astrophysics and cosmology at high redshifts,
is sensitive to the spectrum and nature of the early X-ray sources
\citep{Fialkov:2014, Pacucci:2014, Mirocha:2014, Fialkov:2014b,
  Fialkov:2015, Ewall-Wice:2016}.  In this paper we explore the
redshifted 21-cm signal while constraining the luminosity of the
high-redshift X-ray sources by the observed soft CXRB. The CXRB
constraint yields an upper limit on the X-ray efficiency of each type
of source and thus allows us to estimate the maximal possible effect
of cosmic heating on the 21-cm signal. Such a limit has not been
quoted before, although the effect of strong soft X-ray sources on the
21-cm signal has been studied \citep{Pritchard:2012, Mesinger:2013,
  Pacucci:2014}.  Moreover, we improve over the existing works by
considering the CXRB limits on two new types of X-ray sources, namely,
realistic X-ray binaries and mini-quasars, both with a hard SED.
Given the great current uncertainty about the properties of early
galaxies, this is an essential constraint that will help guide the
search for the radio signal by existing,  planned  and upcoming radio telescopes
designed to probe the 21-cm signal out to $z\sim 30$, such as the
Square Kilometer Array (SKA, \citet{Koopmans:2015}), The Hydrogen
Epoch of Reionization Array (HERA)\footnote{http://reionization.org/},
Large Aperture Experiment to Detect the Dark Age (LEDA,
\citet{Bernardi:2016}), the Experiment to Detect the Global EoR Step
(EDGES, \citet{Bowman:2010}), the Dark Ages Radio Explorer (DARE,
\citet{Burns:2012}), the Shaped Antenna measurement of the background
RAdio Spectrum (SARAS, \citet{Patra:2013}), the SCI-HI experiment
\citep{Voytek:2014}, the Donald C. Backer Precision Array for Probing
the Epoch of Reionization (PAPER, \citep{Parsons:2014}), Giant
Metrewave Radio Telescope (GMRT, \citet{Paciga:2013}), the Murchison
Widefield Array (MWA, \citet{Bowman:2013}), the LOw-Frequency ARray
(LOFAR, \citet{Haarlem:2013}), and the New Extension in Nan\c{c}ay
Upgrading LOFAR (NenuFAR, \citet{NenuFAR}).

The paper is organized as follows: in Section \ref{Sec:sim} we
describe the simulation and model assumptions used in this work; in
Section \ref{sec:CXRB} we (1) use the observed unresolved CXRB to
  establish the upper bound on the X-ray efficiency of high-redshift
  sources, and (2) apply the existing upper limits on the 21-cm power
  spectrum to set the lower limit on the X-ray efficiency; in Section
\ref{Sec:Res} we present our results for the thermal history, partial
X-ray ionization, and the 21-cm signal; finally, we summarize and
conclude in Section \ref{Sec:sum}. Throughout this paper we use
cosmological parameters as measured by the Planck Collaboration
\citep{Planck:2013}.


\section{Simulated Universe}
\label{Sec:sim}

We simulated large cosmological volumes of 384$^3$ Mpc$^3$ of the high
redshift Universe using a hybrid simulation, first introduced by
\citet{Visbal:2012} and described in detail by \citet{Fialkov:2014};
we describe it briefly here. Using the known statistical
  properties of the initial density field, we generated a random
  realization of the initial overdensity (with periodic boundary
  conditions) and the supersonic relative velocities between the gas
  and dark matter \citep{Tseliakhovich:2010, Tseliakhovich:2011,
    Fialkov:2014c} in a cubic volume. Given the large-scale density
  distribution, we then computed the gas fraction in star-forming
  halos in each cell as a function of time. In our simulation the star
  formation rate, found following the extended Press-Schechter
  formalism \citep{Barkana:2004}, is modified by the large-scale
  density fluctuations and the supersonic relative velocities.
  Throughout the simulation we assume Population II stars with
  standard spectra [from \citet{Barkana:2005, Leitherer:1999}] and a
  star formation efficiency of 5\%.  We also account for the effect of
  photoheating feedback on the amount of gas available for star
  formation \citep{Cohen:2015}. Next, we use the stellar distribution
  to determine the X-ray heating rate, ionization, free electron
  fraction and the intensity of the Ly-$\alpha$ background in each
  cell. To this end, we first smooth the stellar density field at each
  redshift in shells around each cell. We assume the flux of X-ray and
  Ly-$\alpha$ photons emitted from each shell to be proportional to
  the star formation rate, which is in turn proportional to the time
  derivative of the amount of gas in star forming halos. We then
  compute the heating and ionization rates as well as the intensity of
  the Ly-$\alpha$ background by integrating over all the shells seen
  by each cell [using rates from \citet{Furlanetto:2010}]. In this
  integral, the contribution of each cell to the rate at a given
  central cell is computed at the time-delayed redshift as seen by the
  central cell. Given the X-ray heating and ionization rates versus
  redshift at each cell, we integrated to get the gas temperature and
  free electron fraction as a function of time. To include UV
  reionization, we set each cell to be fully reionized if some sphere
  around it contains enough ionizing photons to self-reionize
  \citep{Furlanetto:2004}. With this simulation we followed the
  thermal history of the Universe and made predictions for the
  observable soft band CXRB and the 21-cm signal from a wide range of
  redshifts, $z = 6-40$, as discussed in Sections \ref{sec:CXRB} and
  \ref{Sec:Res}. 

This tool, based on a combination of numerical simulation and
analytical calculations has enough flexibility to explore various
parameters of the unconstrained high redshift environment such as the
X-ray efficiency ($f_X$) and SED of the first heating sources, the
minimal mass of star forming halos and the reionization history. To
probe the parameter space we consider two cases in which the IGM is
reionized by UV photons emitted by early galaxies. In the ``late
reionization'' case the Universe is fully reionized by $z_{\rm re} =
6.2$, with the value of the electron optical depth falling in the
range $\tau = 0.059-0.074$ which is within $1-2\sigma$ of the latest
Planck result, $\tau = 0.058\pm 0.012$, \citep{Planck:2016}; while in
the second case, ``early reionization'', $z_{\rm re} =8.5$ and $\tau$
falls within $\sim 3\sigma$ of the Planck measurement.  Late
  versus early reionization scenarios are obtained by varying the
  ionizing efficiency of sources in each model. When we normalize the
X-ray emission (see below), we also use the redshift of full
reionization as the cut-off time for this normalization; in this way,
the two cases with different values of $z_{\rm re}$ also serve to
probe which redshifts are really constrained by the X-ray
background. Next, we consider two possible cases for the typical
galactic halo mass: including star formation in halos down to the
lowest mass that allows for efficient atomic cooling, which we term
the ``Atomic cooling'' case, or adopting a minimum halo mass for star
formation that is ten times larger (allowing for the possibility of
strong feedback in small halos), which we refer to as ``Massive
halos''.  Finally, we consider three types of X-ray sources: X-ray
binaries that have a hard SED \citep{Fragos:2013, Fialkov:2014},
sources with a soft power-law SED of spectral index $\alpha_S = 0.5$
[where the luminosity $L$ follows $dL/d\log\nu \propto
  \nu^{-\alpha_S}$] \citep{Furlanetto:2006b}, and mini-quasars, i.e.,
central black holes in early star-forming galaxies, which we discuss
in greater detail in Section \ref{Sec:MQ}.

The efficiency of hot gas or X-ray binaries is defined via the
relation between the bolometric X-ray luminosity $L_X$ and the star
formation rate (SFR) of the galaxy
\begin{equation}
  \frac{L_X}{SFR} = 3\times10^{40} f_X~ \rm{erg\,s}^{-1}
  \rm{M}_{\odot}^{-1}yr\ ,
\label{eq:LS2}
\end{equation}
where the standard value of the X-ray efficiency is $f_X = 1$
\citep{Furlanetto:2006, Fragos:2013}. This relation is based on
  observations of nearby starburst galaxies and XRBs
  \citep{Grimm:2003, Gilfanov:2004, Mineo:2012a,Mineo:2012b}, and the
  standard normalization for XRBs (with $f_X = 1$) includes an
  order-of-magnitude increase in this ratio at the low metallicity
  expected for high-redshift galaxies \citep{Fragos:2013}.

While the numerical factor in Eq.(\ref{eq:LS2}) was derived for
  X-ray binaries over the range $0.2-95$ keV \citep{Fragos:2013}, in
  this paper we chose a common normalization for soft and hard sources
  for simplicity. For easier comparison with other work in the
  literature, we list here the luminosities (for $f_X=1$) in the
  $0.2-30$ keV and $0.5-8$ keV bands, for both the hard and the soft
  SEDs: $L_{0.2-30}^{\rm soft}/SFR = 3\times 10^{40}$,
  $L_{0.2-30}^{\rm hard}/SFR = 2.8\times 10^{40}$; $L_{0.5-8}^{\rm
    soft}/SFR = 1.6\times 10^{40}$, $L_{0.5-8}^{\rm hard}/SFR =
  2.4\times 10^{40}$. Note that in our late EoR scenario most of the
  X-rays that contribute to the CXRB originate near $z_{\rm re} =
  6.2$, in the rest-frame frequency range of $3.6-14.4$ keV where the
  corresponding luminosities are $L_{3.6-14.4}^{\rm soft}/SFR =
  3.9\times 10^{39}$ and $L_{3.6-14.4}^{\rm hard}/SFR = 1.2\times
  10^{40}$. In the early EoR scenario, i.e., $z_{\rm re} = 8.5$, {\it
    Chandra's} soft band probes the rest-frame $4.7-19$ keV range
  where the luminosities amount to $L_{4.7-19}^{\rm soft}/SFR =
  3.4\times 10^{39}$ and $L_{4.7-19}^{\rm hard}/SFR = 8.7\times
  10^{39}$.

\subsection{Mini-quasars}
\label{Sec:MQ}
High redshift black holes are expected to be much lighter than the
ones observed today in Milky Way size halos, $\sim 10^2- 10^4
~M_\odot$ compared to $\sim 10^6- 10^9 ~M_\odot$. Until recently, the
early population of high redshift black holes with such small masses
was considered purely speculative. However, the latest observations
find black holes with masses in the $10^2- 10^6 ~M_\odot$ range in
dwarf metal poor galaxies which may resemble the high-redshift
environment.  In particular, \citet{Moran:2014} used the Sloan Digital
Sky Survey to find 28 AGN in nearby low-mass, low-luminosity dwarf
galaxies and estimated the minimum black-hole mass to fall mainly in
the $10^3- 10^4 ~M_\odot$ range, showing that AGN in dwarf galaxies
are not as rare as previously thought; \citet{Marleau:2014} studied
the infrared signature of active nuclei in nearby dwarf galaxies and
found black hole masses of $10^2- 10^6 ~M_\odot$; while
\citet{Lemons:2015} found that relatively many hard X-ray sources in
dwarf galaxies are ultra-luminous. The abundance and brightness of
these sources suggest that mini-quasars and high redshift AGN may have
had a stronger impact on reionization than what is usually assumed.

Here we use the internal feedback model for the black hole mass
\citep{Wyithe:2003} and follow the discussion in \citet{Fialkov:2014}
to model the population of high redshift mini-quasars and their impact
on the heating and reionization of the Universe. 
Thus, we find the ratio of the average X-ray luminosity of the central
mini-quasar to that of XRBs in the same halo to be
\begin{equation}
  \frac{L_{MQ}}{L_{XRB}} \sim 0.1 \left(\frac{0.05 f_{X}^{MQ}}{f_X f_\star}\right)
  \left(\frac{M_h}{10^8~M_\odot}\right)^{2/3}\frac{1+z}{10}\ ,
\label{eq:LS}
\end{equation}
where we assumed, as in the case of X-ray binaries, that on
  average 25\% of the mini-quasar X-rays in the relevant wavelengths
  are not absorbed in the interstellar medium of the host galaxy. Here
  we needed this non-absorbed fraction in order to compare between
  XRBs (for whom ISM absorption is implicitly included in the
  spectrum, which is based on observations) to mini-quasars (for which
  we estimated the intrinsic emitted energy using a theoretical
  model). We note that, by using the XRB spectrum, we assume (like
  \citet{Fragos:2013}) that this absorption at high redshifts is
  similar to that at the low redshifts of the observations. In the
  ratio in Eq.~(\ref{eq:LS2}), $f_{X}^{MQ}$ parameterizes the
uncertainty in the X-ray efficiency of mini-quasars with $ f_{X}^{MQ}
=1$ being its standard value. For simplicity, in the rest of this
paper we omit the superscript MQ when talking about the X-ray
efficiency of mini-quasars. Using Eq.~(\ref{eq:LS2}) we can express
the X-ray luminosity of a mini-quasar formed in a halo of mass $M_h$
in terms of the star formation rate in the same halo.

Compared to XRBs, the X-ray luminosity of a mini-quasar is weighted by
an additional factor of $\left(M_h/10^8 ~M_\odot\right)^{2/3}$.
Because of this additional factor, the contribution of mini-quasars to
the X-ray background is expected to be negligible compared to that of
XRBs at high redshifts ($z\gg8$) where the typical halo mass is small,
but it becomes important at lower redshifts ($z\ll8$) at which larger
halos form bigger black holes. When mini-quasar emission is
  modelled as a multi-colour \citet{Shakura:1973} accretion disc,
  these sources are expected to have hard SEDs that are similar to
  those of XRBs, due to the weak dependence of the X-ray spectrum on
  the black hole mass \citep{Tanaka:2012}. Therefore, we use the same
  SED for XRBs and mini-quasars in our calculation.

\section{Limits on the X-ray efficiency}
\label{sec:CXRB}

In this work we rely on the observed intensity of the unresolved CXRB
in the $0.5-2$ keV band as measured by {\it Chandra}
\citep{Lehmer:2012} to set the maximum possible X-ray efficiency for
each type of the considered X-ray sources.  Following the model
developed by \citet{Cappelluti:2012}, we assume the extra-galactic
contribution amounts to $2.51\times 10^{-13}~\rm{erg
  ~cm}^{-2}\rm{s}^{-1}\rm{deg}^{-2}$ and is provided solely by the
pre-reionization sources. This normalization gives the maximal value
of X-ray efficiency of the high-redshift heating sources that does
not violate the observed soft band CXRB.

The values of $f_X$ that we obtain by comparing the simulated
background to the observed one are listed in the third column of
Table~1 for each case of the reionization history and SED specified in
the first two columns of the Table. We find that in order to produced
the detected unresolved X-ray background, $f_X$ must be 1--2 orders of
magnitude larger than the standard value, $f_X\sim1$, normally assumed
in the literature for the efficiency of the high-redshift sources,
although even higher values of $f_X$ have been considered
\citep{Christian:2013, Mesinger:2013, Pacucci:2014}.
  
We find that hard sources, and in particular mini-quasars, are more
efficient in producing the CXRB in the soft band, in that they require
an X-ray efficiency that is a factor of a few lower than that of soft
sources.  This is because most of the energy emitted by soft X-ray
sources redshifts to frequencies much lower than the observed band.
Therefore, a high efficiency is needed in that case in order to
increase the number of photons that eventually redshift into the
relevant energy range. On the other hand, in the case of the hard
X-ray sources (XRBs and mini-quasars) the SED peaks at $\sim 3$ keV
and, at least for sources at $z\sim 6$ existing just before the end of
reionization in our late reionization scenario, the observed $0.5-2$
keV band probes the energy range around the peak of the SED, and a
lower efficiency is sufficient to produce the observed CXRB.

\begin{table*}
\begin{center}
\begin{tabular}{|l |l  | l |l| l| l| l|l| l|l|l|l|l|l|l|}
  \hline
  Model &  SED & $f_{X}$ &$z_{h}$ & $x_{\rm ion}^X$ & $\tau_{\rm UV}$ & $\tau_{\rm tot}$&   $T_{21}^{\,\rm min}$\,[mK] & $z_{\rm min}$ &  $T_{21}^{\, \rm max}$\,[mK]& $z_{\rm max}$\\
  \hline
   Massive &    Hard  &14.7& 13.5 & 7.6\%&0.0597 & 0.0609 &      -85.3 &  17.0 & 20.8 & 10.7
   \\
   Late    EoR     & Soft & 41.4&  17.0 & 27.3\% &  0.0601 & 0.0688 &    -29.7 &  18.8 & 26.2 & 12.8 
   \\
   $z_{\rm re}^{\rm UV} = 6.2$            & MQ & 12.1 & 13.0 & 6.4\%   &0.0597 & 0.0606  &   -99.7  & 16.7  & 19.6 & 10.4
   \\

\hline  
   Massive               &   Hard  & 79.2&  15.7 & 9.3\%  & 0.0831 &0.0850 &    -42.7& 18.5  & 22.7 & 12.8
   \\
  Early  EoR          & Soft  &187.7 & 18.8 & 34.5\% & 0.0833&  0.0934&    -15.1 &   20.3   & 26.3 & 14.3 
  \\		
   $z_{\rm re}^{\rm UV} = 8.5$      	   &   MQ &87.9 & 15.4  & 8.6\% & 0.0831& 0.0847  & -47.5  & 18.2  & 22.2 & 12.5
   \\
        
\hline     
   Atomic &    Hard & 10.8&16.2&  10.5\%  & 0.0739 & 0.0756&    -107.4& 20.6 & 23.0 & 13.0
    \\   
  Late EoR & Soft & 29.5 &20.1 &  40.0\% &  0.0746& 0.0859&    -48.8& 22.7  & 29.7 & 15.6
  \\  
 $z_{\rm re}^{\rm UV} = 6.2$           &  MQ & 11.4  &14.0 & 7.5\% &  0.0738& 0.0747 &    -147.4&  19.3 & 17.7 & 11.6
 \\ 
   
\hline    
  Atomic        &   Hard & 44.4 &  18.5 & 13.5\% & 0.0957& 0.0990&   -68.8 &  21.9  & 25.9 & 15.2
  \\          
  Early  EoR        & Soft & 102.4 & 21.9 & 39.1\%& 0.0961& 0.1111  &   -29.2  &   24.1  & 30.4 & 17.3
  \\         
       $z_{\rm re}^{\rm UV} = 8.5$             &  MQ& 74.4 & 16.9& 10.5\% &0.0956 & 0.0977 &   -99.0& 20.9  & 22.5 & 14.0
       \\                
  \hline
\end{tabular}
\caption{For each scenario of structure formation and reionization
  (column 1; 4 total scenarios) we list the SED (column 2), $f_X$
  (column 3) set so that the high-redshift sources produce the entire
  unresolved CXRB in the 0.5--2 keV band of $2.51\times
  10^{-13}~\rm{erg ~cm}^{-2}\rm{s}^{-1}\rm{deg}^{-2}$
  \citep{Cappelluti:2012}, the heating transition redshift $z_h$
  (column 4), the X-ray ionization fraction at the end of the UV
  reionization (column 5), the UV contribution to the optical depth
  (column 6) and the total optical depth $\tau_{\rm tot}$ due to the
  UV and X-ray ionization (column 7), the minimal temperature of the
  global 21-cm signal ($T_{21}^{\, \rm min}$, column 8), and the
  redshift at which it occurs ($z_{\rm min}$, column 9),  the
    maximal temperature of the global 21-cm signal in emission
    ($T_{21}^{\, \rm max}$, column 10), and the redshift at which it
    occurs ($z_{\rm max}$, column 11).  }
\end{center}
\label{Tab:1}
\end{table*} 
   
\begin{figure}
\centering
\includegraphics[width=3.4in]{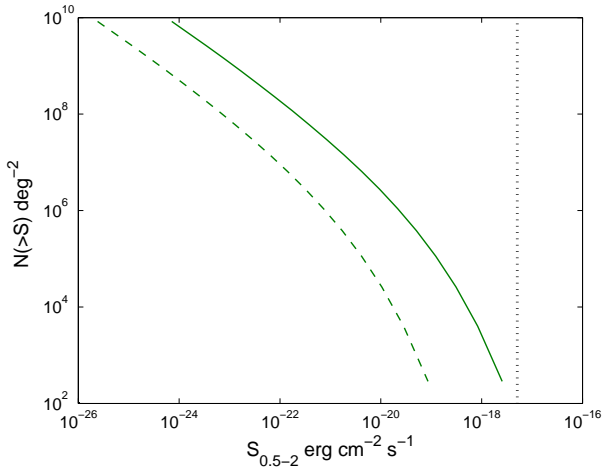}
\caption{Cumulative X-ray number counts at $z = 6.2$ for the soft band
  with atomic cooling halos and a soft SED, for $f_X = 29.5$ (solid)
  or $f_X = 1$ (dashed). The vertical dotted line shows the detection
  limit of {\it Chandra} in the soft band.  }
\label{fig:01}
\end{figure}

Even though the maximum X-ray efficiencies that we find are relatively
high, the population of pre-reionization sources with such
efficiencies is still very dim and cannot be resolved into point
sources by {\it Chandra}.  In Figure~\ref{fig:01} we show the
cumulative number counts expected for such a population for the case
of the high-redshift X-ray sources with a soft SED, with $f_{X}$ =
29.5 (and $f_X =$ 1 for comparison).  As we see from the Figure the
fluxes are well below the detection limit of the satellite, $\sim
5.1\times 10^{-18}~\rm{erg ~cm}^{-2}\rm{s}^{-1}$, and, therefore, it
is correct to consider these sources as unresolved ones, as we do in
this work.

Another trend in the normalized X-ray efficiencies $f_X$ in Table~1 is
that the early reionization cases correspond to a higher $f_X$. This is
not due to reionization itself but since (as noted in
Section~\ref{Sec:sim}) we use the end of reionization as a convenient
cutoff redshift for the X-ray source population, in order to probe how
various redshifts are constrained. Thus, in the case of early
reionization, when the X-ray sources that we are considering are
assumed to be cut off below $z=8.5$, we require a very high
normalization at $z>8.5$ in order to explain the observed CXRB.
Another way to express this is that sources at higher redshifts are
less strongly constrained, i.e., a higher maximal $f_X$ is allowed for
them without violating the observed CXRB.  The $f_X$ values shown in
Table~1 can be seen as upper limits on a hypothetical high-redshift
population of X-ray sources that formed down to $z=6.2$ or $z=8.5$,
and may have had a much higher X-ray efficiency than is typical for
low-redshift galaxies. In the case of Massive halos, there are fewer
galaxies at any given redshift than in the Atomic case, so the X-ray
efficiency must be higher in order to explain the same, fixed, 
level of the observed CXRB.

\begin{figure}
\centering
\includegraphics[width=3.4in]{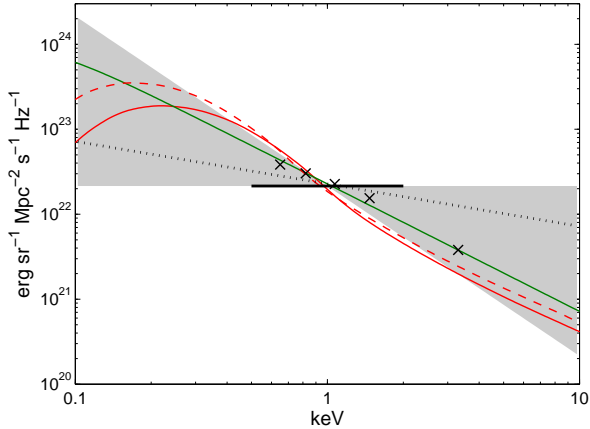}
\caption{Extra-galactic contribution to the unresolved X-ray
  background in the $0.5-2$ keV band versus the energy of the photon.
  Assuming atomic cooling halos, we show the cases of a hard spectrum
  for early (red dashed) or late (red solid) reionization, and a soft
  SED for early (green dashed) or late (green solid) reionization; the
  two soft SED cases lie on top of each other. We calibrate the CXRB
  in the soft band $0.5-2$ keV to the value of $2.51\times
  10^{-13}~\rm{erg ~cm}^{-2}\rm{s}^{-1}\rm{deg}^{-2}$
  \citep{Cappelluti:2012}, showing the average intensity with the
  black horizontal bar in the plot. We also show the shape of the
  ``absolute'' unresolved soft extragalactic X-ray background from
  \citet{McCammon:2002} (crosses) normalized to the same value. The
  shaded area spans the range of slopes considered by
  \citet{Dijkstra:2012} for the CXRB (between $\alpha = 0$ and $\alpha
  = 2$), and the black dotted curve has the best-fit spectral index
  from {\it Chandra} data \citep{Hickox:2006}, $\alpha = 0.5$.}
\label{fig:1}
\end{figure}

The shape of the unresolved CXRB in the $0.5-2$ keV band shown in
Figure \ref{fig:1} is poorly constrained by observations, and only a
general trend can be extracted from the available data at the
moment. The best fit to {\it Chandra} data proposed by
\citet{Hickox:2006} is a power law with a spectral index of $\alpha =
0.5$ (but note that the amplitude of X-ray number counts, not the
slope, was a free parameter in this fit).  Another constraint on the
shape of the CXRB comes from the data collected by the {\it
  R{\"o}ntgenstrahlen} (ROSAT) satellite \citep{Hasinger:1993};
\citet{McCammon:2002} reported the spectral structure of the
unresolved background measured in the $\sim 0.65-3.3$ keV range. These
data were extracted after subtracting the local thermal component, the
contribution of oxygen lines from the ROSAT data, and accounting for
the AGN directly observed by the deep surveys with the
satellite. However, due to a considerable discrepancy in the
normalization of the CXRB measured by {\it Chandra} and ROSAT, which
most likely arises from cosmic variance and systematics, it is unclear
if the residual found in ROSAT data can be fully attributed to the
cosmic background. Therefore, we only show the unresolved CXRB
measured by ROSAT (black crosses in Figure~\ref{fig:1}) with its
normalization re-scaled to the value of $2.51\times 10^{-13}~\rm{erg
  ~cm}^{-2}\rm{s}^{-1}\rm{deg}^{-2}$ in the $0.5-2$ keV band.

Despite the discrepancy in the normalization and large uncertainties
in the ROSAT data, it is still interesting to compare the shape of the
CXRB found by \citet{McCammon:2002} to what is extracted from our
simulations (Figure \ref{fig:1}). This shape is in good agreement with
the CXRB which we get using our soft SED while our hard SED yields a
slightly steeper frequency dependence; the fit to {\it Chandra} data
is somewhat softer than predicted by our models.  Because
observational constraints on the slope are so poor and inconclusive, a
wide variety of possible SEDs are considered in the literature when
modeling high-redshift populations. For instance,
\citet{Dijkstra:2012} explored a wide range of spectral indices,
$\alpha = 0-2$, of high-redshift galaxy SEDs when trying to reproduce
the observed CXRB (this range is marked as the shaded area in Figure
\ref{fig:1}). Both our cases (soft and hard SEDs) yield a CXRB with
slopes which (approximately) fall within this range.  We note that our
normalization at $0.5-2$ keV is consistent with \citet{moretti}, but
their data suggest a somewhat harder spectrum towards higher energies
than the range shown in Figure~\ref{fig:1}.

A lower limit on the X-ray efficiency can in some cases be extracted
from the upper limit on the 21-cm power spectrum \citep{Pober:2015,
  Ali:2015}.  As we will see in the next Section, the hotter is the
IGM, the smaller is the amplitude of the 21-cm signal seen against the
CMB.  Consequently, a very cold IGM results in fluctuations of
  large amplitude that can be ruled out by existing observations.
  \citet{Parsons:2014} were the first to place limits on X-ray heating
  with the 21-cm power spectrum using the data collected by the PAPER
  experiment. Current upper limit on the 21-cm power spectrum of $\sim
  500$ mK$^2$ at $z = 8.4$ in the range of wavenumbers $ k =
  0.15-0.5$~h~Mpc$^{-1}$ \citep{Pober:2015, Ali:2015} yields $f_X
  \gtrsim 0.0095$ (0.0023) in the case of a hard (soft) SED for our
  atomic cooling case and $f_X \gtrsim 0.036$ (0.01) in the case of
  massive halos for the late EoR scenario. Our early reionization
  scenario is not constrained by this measurement since the universe
  is fully ionized by $z_{\rm re}=8.5$ and the 21-cm signal at lower
  redshifts vanishes for all heating scenarios.

\section{Results}
\label{Sec:Res}
\subsection{Ionization and Heating}

In this section we present our quantitative results for the possible
consequences of having strong X-ray sources at high redshift as
currently allowed by the CXRB. One obvious result is that X-ray
sources with a high X-ray efficiency heat up the IGM much earlier than
is normally expected. The moment when the gas heats up to the
temperature of the CMB (the cosmic heating transition), of which
redshift we denote $z_h$, happens earlier in the models where $f_X$ is
set to its upper limit compared to the predictions for the $f_X=1$
case. In the case of the soft SED, late reionization and atomic
cooling, this transition happens a $\Delta z_h = 5.3$ earlier, while
in the case of XRBs the advance is $\Delta z_h = 4.2$.  In Table 1 we
list $z_h$ for all the considered cases with the maximum $f_X$, while
Table 2 shows the same cases but for $f_X = 1$.

 It is interesting to see if the saturated heating assumption,
  which is often used in numerical simulations, holds during
  reionization when the CXRB limit is applied. Saturated heating is
  the assumption that the gas is much hotter than the CMB, i.e.,
  $1-T_{CMB}/T_K\sim 1$. If this assumption is valid, then the heating
  history can be ignored when reionization is considered. Considering
  our models with $f_X$ values from Table~1, we find that in the case
  of a hard SED the value of the factor $1-T_{CMB}/T_K$ is far from
  unity at the beginning of the EoR when the Universe is $\sim 5\%$
  ionized.  In particular in the case of late (early) reionization and
  atomic cooling, this factor (when averaged over a simulated box) is
  $0.6$ ($0.59$) at $x_{HI}=95\%$ for X-ray binaries and $-0.36$
  ($-0.1$) for mini-quasars.  In the case of late (early) reionization
  and massive halos, this factor is $0.66$ (0.75) for X-ray binaries,
  and 0.5 (0.7) for mini-quasars.  By the mid-point of UV
  reionization, $1-T_{CMB}/T_K\gtrsim 0.95$ in all the simulated cases
  from Table~1 and, hence, the saturated heating assumption applies
  during the second half of reionization. The CXRB-normalized models
  with a soft SED yield $1-T_{CMB}/T_K\gtrsim 0.97$ already at an
  ionization fraction of $5\%$.
 \begin{table}
\begin{center}
\begin{tabular}{|l |l  | l |l| l| l| l|l| l|l|l|l|l|l|l|}
  \hline
Model &  SED &  $z_{h}$ & $x_{\rm ion}^X$ &  $T_{21}^{\, \rm min}$\,[mK]& $z_{\rm min}$   \\
  \hline
   Massive &   Hard &9.7& 0.7\%  & -167.7 & 14.9
    \\
   Late EoR       &   Soft& 12.2 &2.3\% & -121.3 & 16.1  
   \\
     $z_{\rm re}^{\rm UV} = 6.2$           &  MQ & 9.5 &0.6\% & -174.3 &  14.6
     \\   
\hline  
   Massive                  &   Hard   & 9.6 &  0.2\% & -163.9& 14.9 
    \\
   Early  EoR         & Soft  & 12.1 & 0.5\%& -120.9&  16.1 
     \\		
    	  $z_{\rm re}^{\rm UV} = 8.5$   &    MQ & 9.3 & 0.1\%& -171.7&   14.6
    	  \\
\hline 
   Atomic &    Hard    &12.0 & 1.3\% & -169.6 & 18.2 
    \\   
 Late EoR  &  Soft  & 14.8 & 4.5\%&-134.8 &19.6 
   \\  
 $z_{\rm re}^{\rm UV} = 6.2$        &    MQ   & 10.1 & 0.8\%&-201.3 & 16.6
  \\ 
\hline  
   Atomic        &   Hard   & 12.0 & 0.5\% & -169.6&  18.3
      \\          
    Early  EoR    & Soft   & 14.8 & 1.6\% &-134.7 & 19.7  
     \\         
    $z_{\rm re}^{\rm UV} = 9$           &    MQ  & 10.0 &  0.2\%&-202.9& 16.5 
     \\ 
  \hline
\end{tabular}
\caption{$f_X = 1$ case. For each scenario of structure formation and
  reionization (column 1), we show the SED (column 2), the heating
  transition redshift $z_h$ (column 3), the X-ray ionization fraction
  at the end of the UV reionization (column 4), the minimal
  temperature of the global 21-cm signal ($T_{21}^{\, \rm min}$,
  column 5), and the redshift at which it occurs ($z_{\rm min}$, column
  6).  }
\end{center}
\label{Tab:2}
\end{table} 

Along with heating, X-rays partially ionize the neutral medium,
competing with the UV photons in powering reionization. The efficiency
of X-rays versus UV in ionizing the cosmic gas depends on the values
of the escape fraction and the X-ray efficiency.  Here we keep the
former parameter fixed in each case by fixing the redshift at which
reionization is completed by the UV photons, while varying the latter.
In the standard case of $f_X = 1$, the main source of ionization is UV
photons while X-rays are expected to yield only a small contribution,
ionizing the gas only up to a few percent when reionization is
completed by UV. To emphasize this, we list the level of the partial
ionization by X-rays (within the final remaining neutral regions just
before they are reionized), $x_{\rm ion}^X$, at the end of
reionization powered by UV photons, in the fourth column of Table 2.
The value of $x_{\rm ion}^X$ does not rise above $4.5\%$.  

On the other hand, with $f_X$ set to its maximum by normalization to
the CXRB, the X-ray efficiency is substantially higher and in some
cases the X-rays start competing with UV in the quest for reionizing
the Universe (see the $x_{\rm ion}^X$ column in Table~1).  Partial
ionization by X-rays is especially high in the case of the soft
spectrum due to the extremely high X-ray efficiency in combination
with the rapid, efficient absorption of soft X-rays by neutral gas.
For instance, by the end of the UV reionization it reaches $x_{\rm
  ion}^X \sim 40\%$ in the case of atomic cooling scenario. 
  For XRBs and mini-quasars, $x_{\rm ion}^X$ grows
to 13.5\% and 10.5\%, respectively, by the end of UV reionization (for
the case of atomic cooling and early reionization). It is interesting
to note that while UV photons heat and ionize the gas close to the
source (given their very small mean free path within neutral gas),
X-rays travel larger distances before inputting their energy in the
IGM. As a result, in the extreme cases with high $f_X$, reionization
is much more homogeneous, and may even proceed outside-in
\citep{Mesinger:2013, Majumdar:2015} instead of the conventional
inside-out picture \citep{Barkana:2004}.

\subsection{21-cm signal}
\label{Sec:results}

The brightness temperature of the 21-cm signal, observed with the CMB
as a diffuse background source, is expected to be a three-dimensional
probe of the high-redshift Universe. Current and planned telescopes
are designed to measure the redshift evolution of this signal averaged
over two-dimensional spheres (the global spectrum) as well as its
spatial fluctuations at each given epoch which will allow us to
extract the power spectrum.  The predicted global spectrum contains
information about milestones in the evolution of the Universe and for
a wide range of models exhibits a prominent absorption trough which
single-dish radio telescopes seek to observe. The sky-averaged
brightness temperature reaches its minimal value at the point when
heating sources turn on and begin raising the temperature of the IGM
towards $T_{\rm CMB}$.  The power spectrum at each comoving
wavenumber, $k$, when plotted as a function of redshift, shows a
sequence of peaks driven by various spatially and temporally varying
physical quantities (such as the kinetic gas temperature, fraction of
neutral gas, and intensity of Ly-$\alpha$ radiation) which determine
the intensity of the 21-cm transition.  A generic plot of this type
for the 21-cm power spectrum has three peaks \citep{Barkana:2005,
  Pritchard:2007, Pritchard:2008}: the high-redshift peak, at $z \sim
20 - 30$, is due to Ly-$\alpha$ fluctuations; the mid-redshift peak,
which is due to heating fluctuations, appears at $z \sim 15 - 22$,
 but this peak is not present on scales below the X-ray mean free
  path \citep{Fialkov:2014b}; and the low-redshift peak, at $z \sim
7-10$, is due to ionization fluctuations.  
of the main targets of the present-day and future radio
interferometers.

\begin{figure*}
\centering
\includegraphics[width=3.4in]{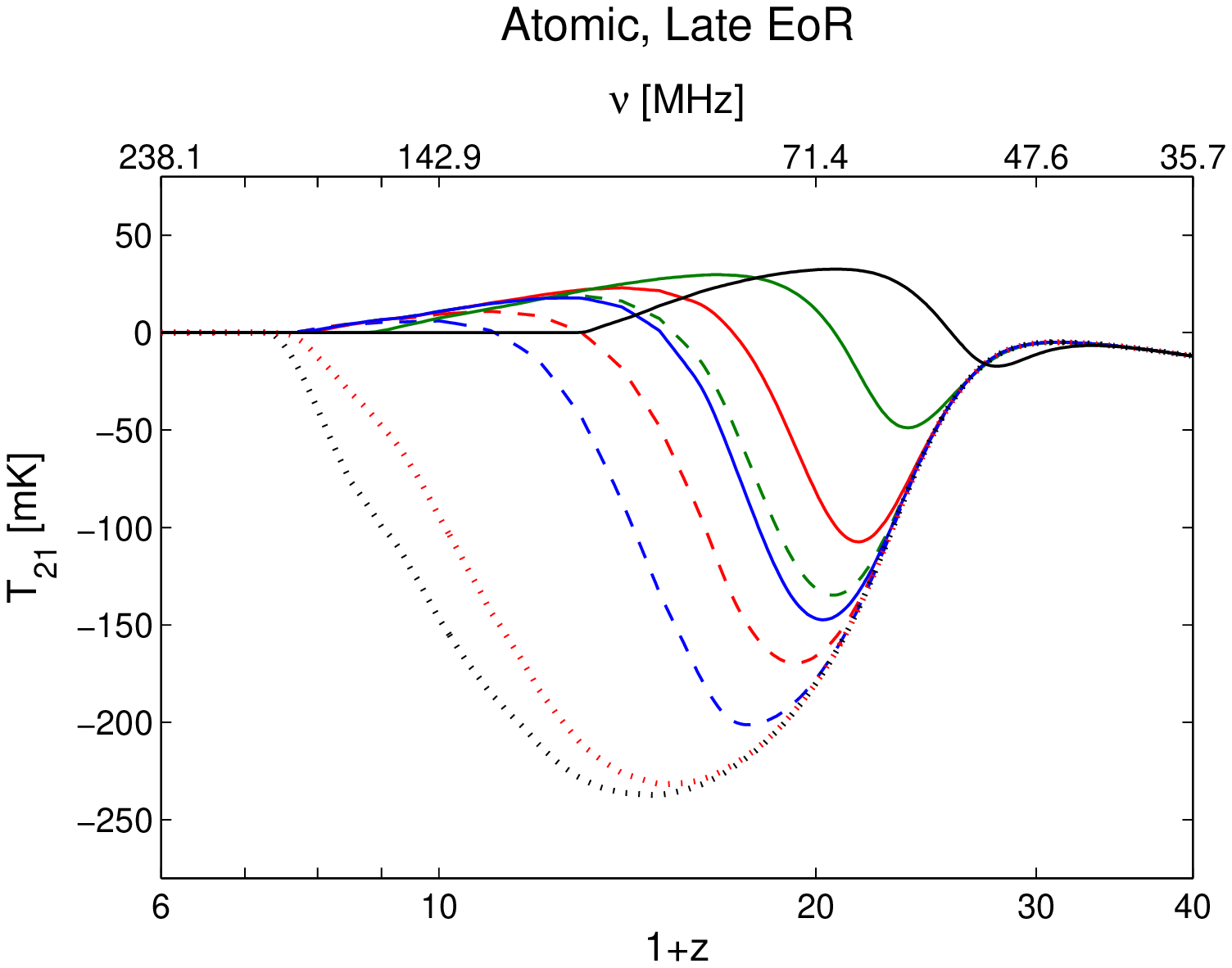}\includegraphics[width=3.4in]{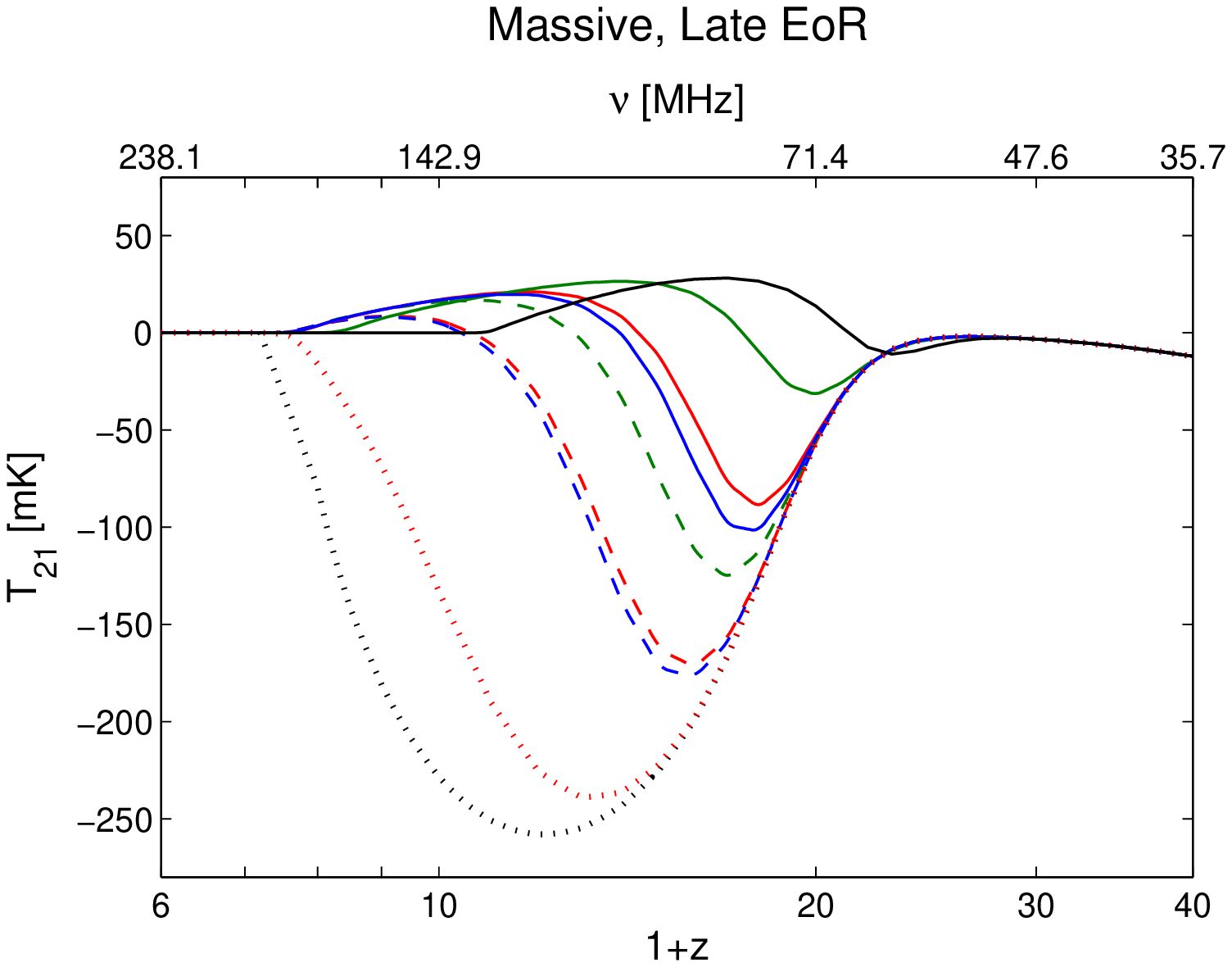}
\includegraphics[width=3.4in]{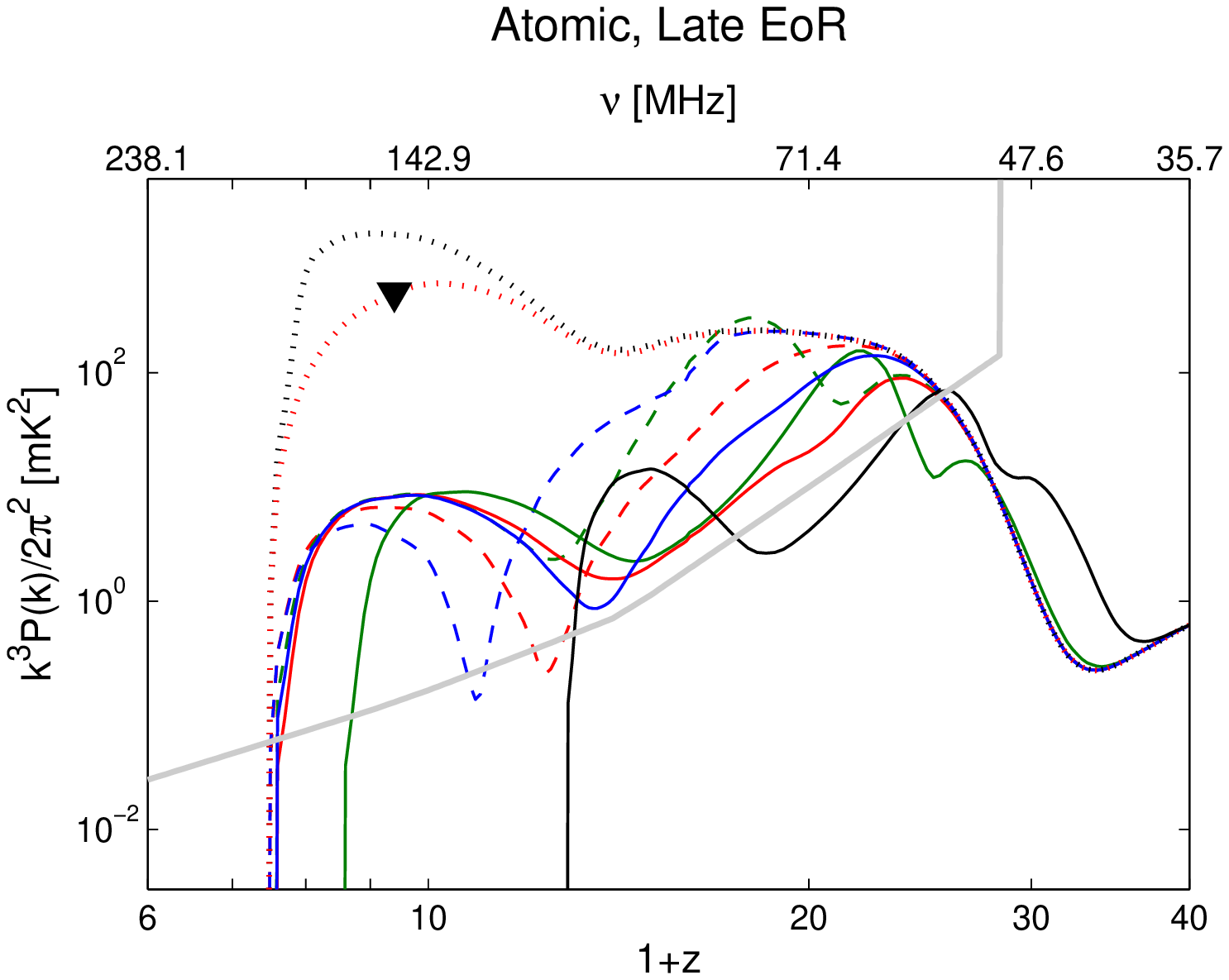}\includegraphics[width=3.4in]{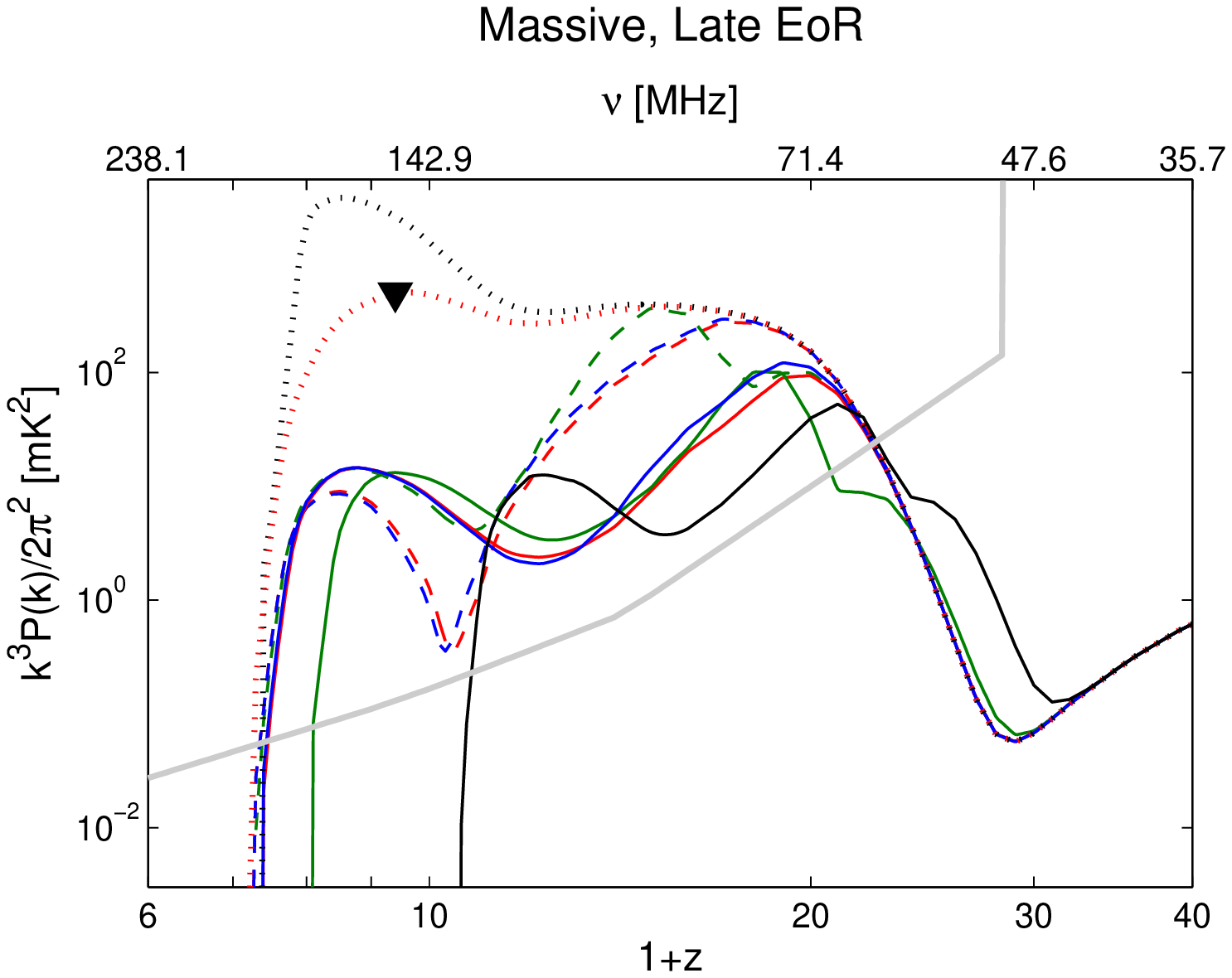}
\caption{Global 21-cm brightness temperature (top) and its power
  spectrum in mK$^2$ units at $k = 0.2$ Mpc$^{-1}$ (bottom) are shown
  for the atomic (left) and massive (right) cases for our late
  reionization case ($z_{\rm re}^{\rm UV} = 6.2$). We show the cases
  of X-ray sources with a soft SED (green), XRBs with a hard SED
  (red), or mini-quasars (blue), each for $f_X = 1$ (dashed) or
  $f_{X}$ normalized to its maximum value based on the CXRB
  (solid). The red dotted line shows the case with a low heating
  efficiency of $f_X = 0.0095$ (0.036) for atomic cooling (massive
  halos) with a hard X-ray SED. The black triangle marks the upper
  limit on the 21-cm power spectrum at $z = 8.4$ \citep{Pober:2015}
   and the black dotted line shows the case of no heating,
    $f_X=0$, which has been ruled out by this observation. In both the atomic and massive cases the black
    solid line shows the case of extreme heating ($f_X = 1000$, soft SED) which
    over-predicts the CXRB. Finally, the thick grey line shows the
    power spectrum of the thermal noise for phase 1 of SKA-low
    assuming a single beam, 1000-hour integration time, and a 10 MHz
    bandwidth.  }
\label{fig:2}
\end{figure*}

\begin{figure*}
\centering
\includegraphics[width=3.4in]{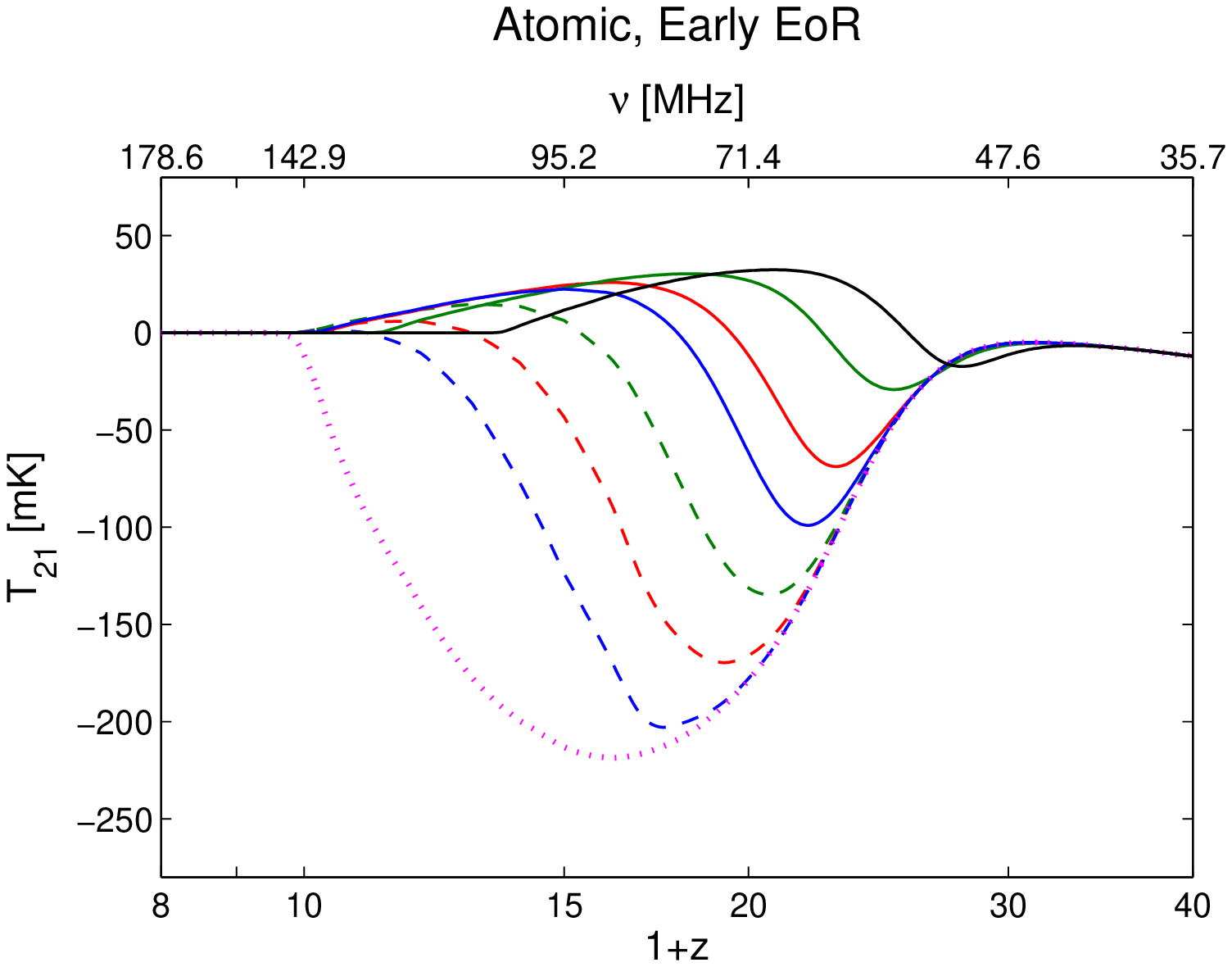}\includegraphics[width=3.4in]{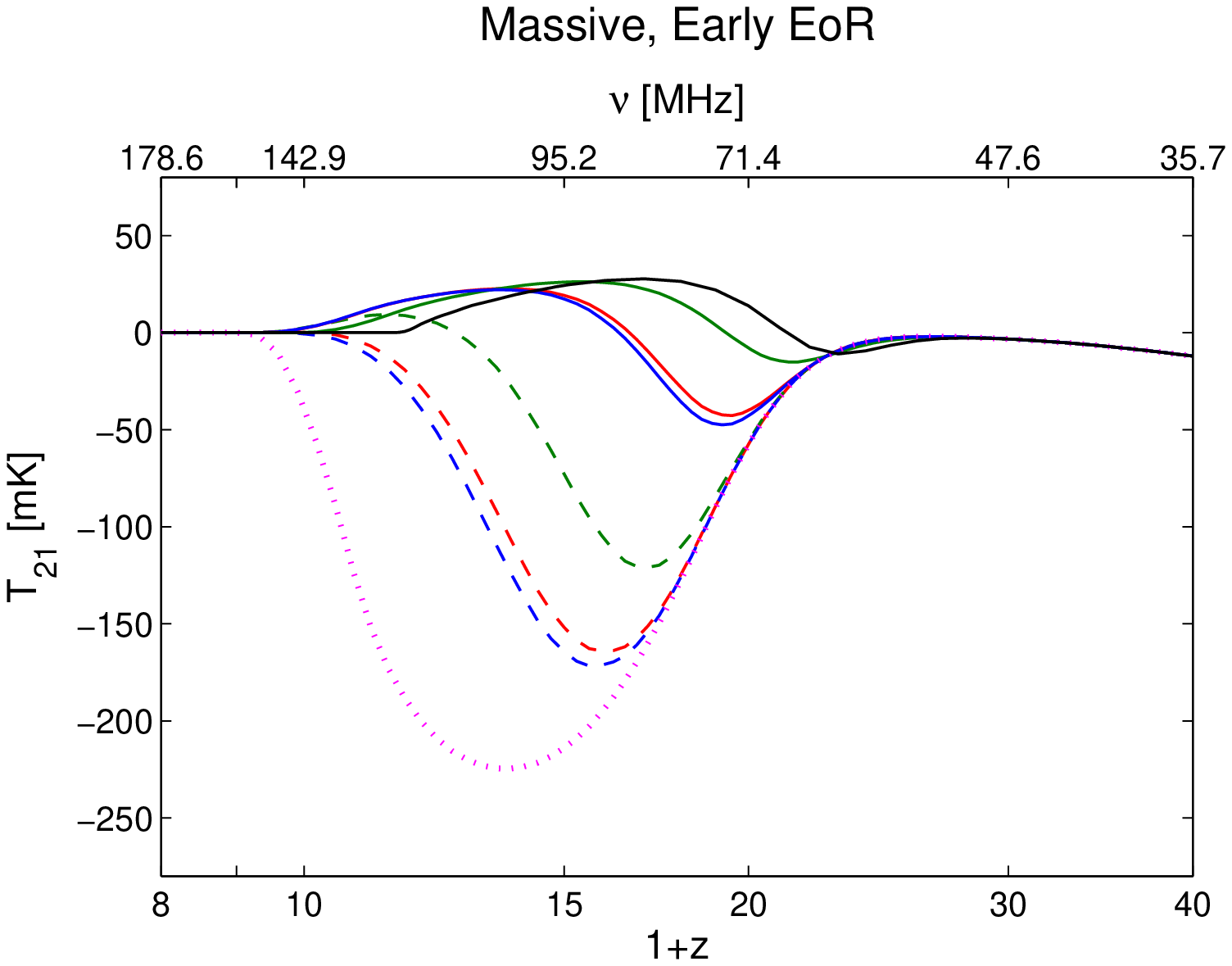}
\includegraphics[width=3.4in]{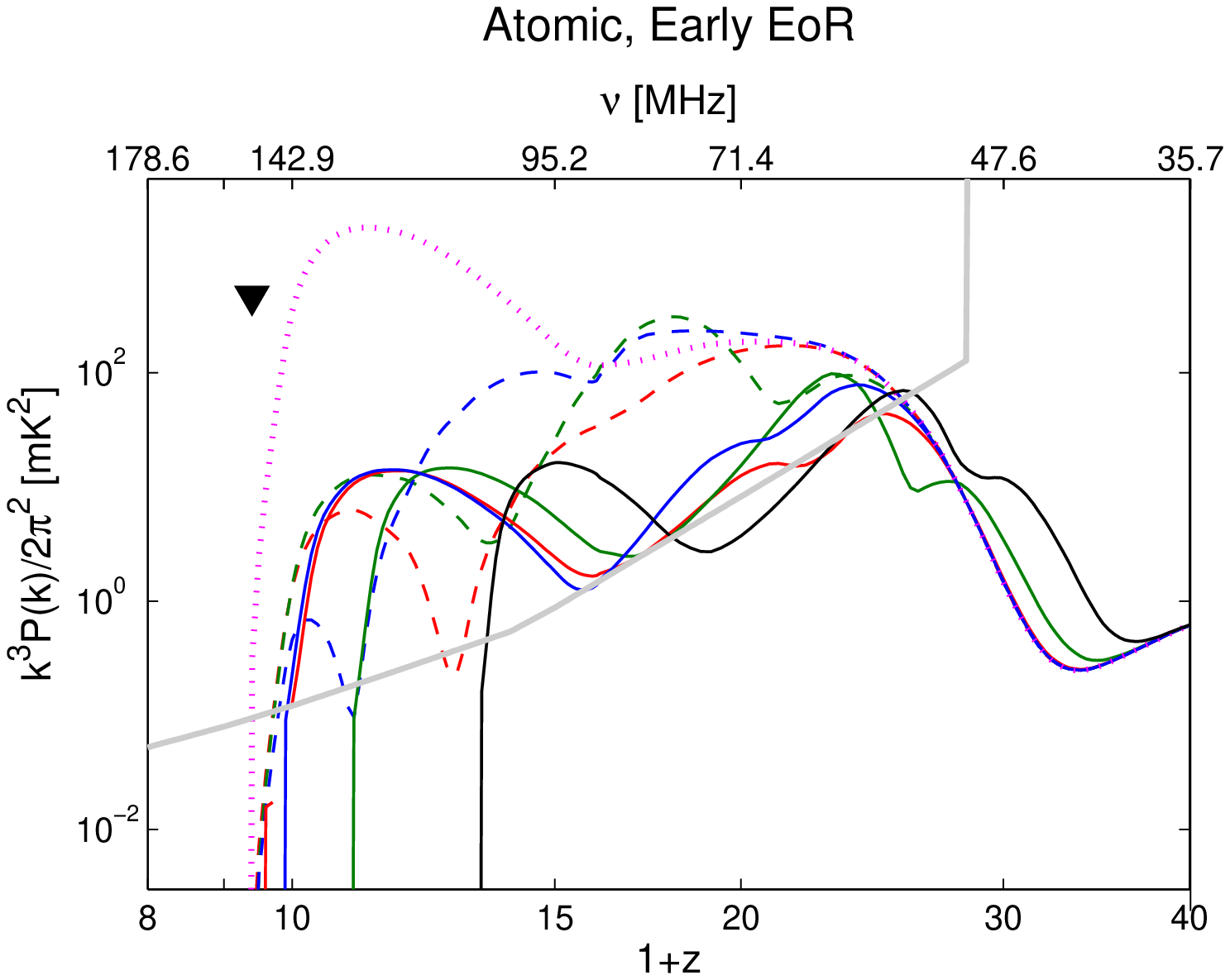}\includegraphics[width=3.4in]{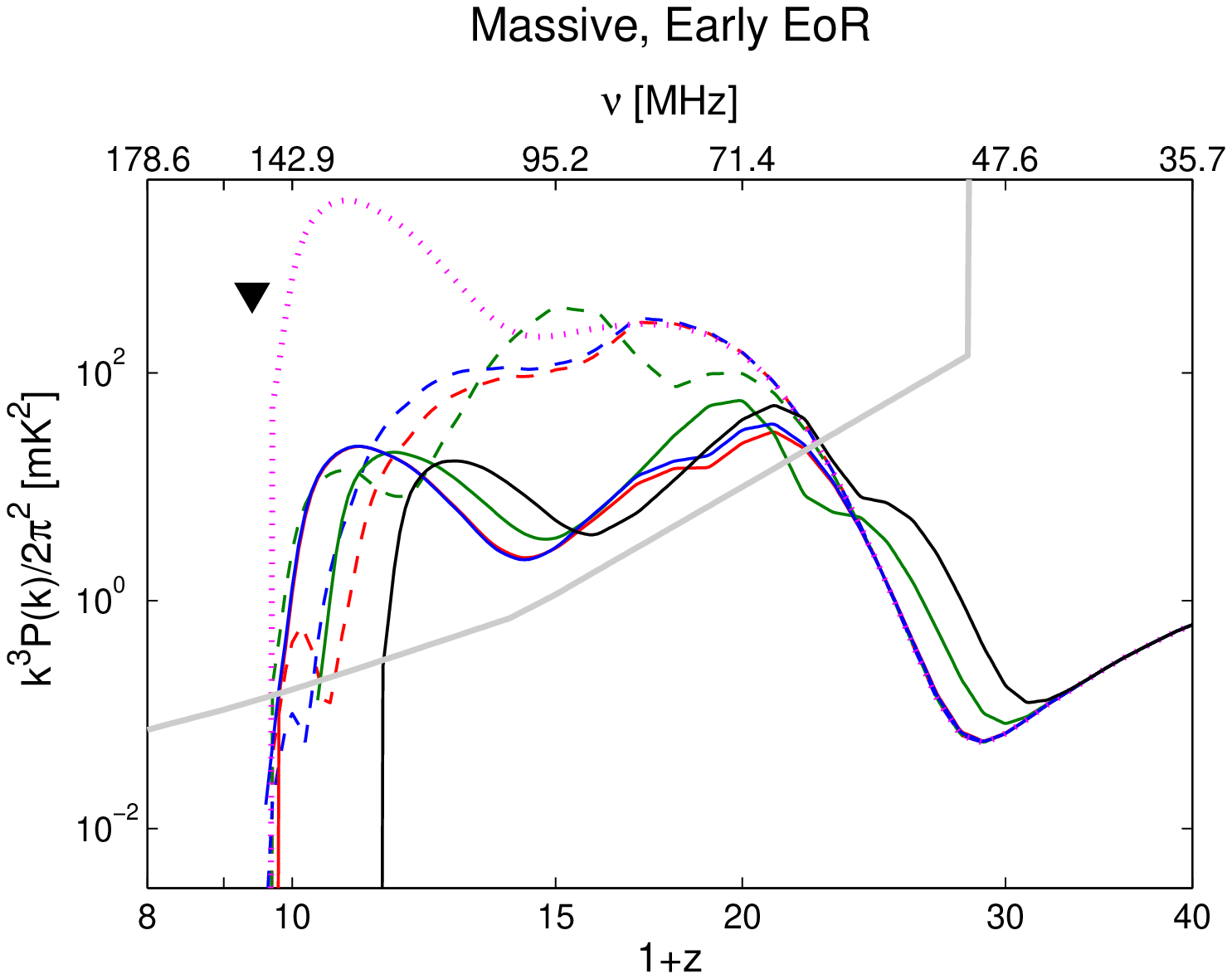}
\caption{Global 21-cm brightness temperature (top) and its power
  spectrum in mK$^2$ units at $k = 0.2$ Mpc$^{-1}$ (bottom) are shown
  for the atomic (left) and massive (right) cases for our early
  reionization case ($z_{\rm re}^{\rm UV} = 8.5$). We show the cases
  of X-ray sources with a soft SED (green), XRBs with a hard SED (red)
  or mini-quasars (blue), each for $f_X = 1$ (dashed) or $f_{X}$
  normalized to its maximum value based on the CXRB (solid). The
  magenta dotted line shows the case with no heating, i.e., $f_X =
  0$.  The black triangle marks the upper limit on the 21-cm power
    spectrum at $z = 8.4$ \citep{Pober:2015}. In both the atomic and massive cases the black
    solid line shows the case of extreme heating ($f_X = 1000$, soft SED) which
    over-predicts the CXRB. Finally, the thick grey line shows the power spectrum of the
    thermal noise for phase 1 of SKA-low assuming a single beam,
    1000-hour integration time, and a 10 MHz bandwidth.  }
\label{fig:3}
\end{figure*}

In Figures \ref{fig:2} (for the late reionization scenario) and
\ref{fig:3} (which assumes early reionization) we plot the global
spectrum of the 21-cm signal and its spherically-averaged power
spectrum, with the X-ray efficiencies normalized to the CXRB as well
as the cases with $f_X=1$, $f_X =0$ and $f_X=1000$  shown for
comparison.  The latter case (which we shown only for the soft SED) clearly over-produces the observed
  CXRB, while the case of $f_X = 0$ and late reionization is excluded
  by the PAPER observations. We do show these cases to demonstrate the
  effect of either extremely strong or absent X-ray heating (the
  excluded cases are shown with black lines in Figures \ref{fig:2} and
  \ref{fig:3}). In addition, for the late reionization scenario and
hard SED shown in Figure \ref{fig:2} we demonstrate the 21-cm signal
with the lowest possible X-ray efficiencies (red dotted lines)
normalized to the current PAPER limits (black triangle) as discussed
at the end of Section \ref{sec:CXRB}. Finally, we plot the expected
thermal noise power spectrum of phase 1 of the SKA assuming a single
beam, an integration time of 1000 hours, and a 10 MHz bandwidth. 
  In the case of global 21-cm experiments, in principle, short
  integration times (minutes) suffice to detect the signal with a
  single dish \citep{Shaver}. In practice, though, the need for
  accurate calibration and precise removal of the spectrally smooth
  foreground make these measurements quite difficult, especially given
  the complex coupling of the errors at different
  frequencies. Quantitative analyses show that if calibration can
  remove any sharp frequency response, then it is possible to overcome
  the issues of a smooth foreground and thermal noise, and global
  signals such as those we predict in this paper should be detectable
  \citep{Pritchard:2010,Liu:2013,Morandi,Bernardi:2016}.

 We find that the cases of enhanced heating of the IGM strongly affect
 the expected global 21-cm signal. In particular, because the gas does
 not have sufficient time to cool down prior to being heated by
 X-rays, the absorption trough becomes shallower and thus harder to
 observe,  requiring a higher precision of calibration and longer
   integration times. For instance, in the case of late reionization
 and atomic cooling, the minimal brightness temperature is reduced by
 $\sim 64\%$ for a soft SED and $\sim 37\%$ for a hard SED. In
 addition, the absorption trough is shifted to lower frequencies: from
 70 MHz to 60 MHz ($\Delta z \sim 3.1$) in the case of the soft SED,
 from 74 MHz to 66 MHz ($\Delta z \sim 2.4$).  Finally, with the
   enhanced heating the emission signal is much stronger and the peak
   shifts to higher redshifts; for each model the maximal possible
   value of the emission signal is realized when the normalization is
   set so as to produce the full unresolved CXRB. In particular, in
   the case of late reionization and atomic cooling, the maximal
   brightness temperature is boosted by a factor of $\sim 2$ ($\sim
   1.6$) for XRBs (hot gas). See Tables 1 and 2 for more
   information. 

As discussed above, X-rays speed up reionization and the intensity of
the 21-cm signal drops faster than otherwise expected towards the end
of the reionization era,  which is most noticeable in the cases
  with soft X-rays and maximal efficiency (set to the full CXRB), as
  well as in the excluded case of extreme heating (with $f_X = 1000$)
  where reionization ends considerably earlier than in other models.
In addition, Ly-$\alpha$ photons generated by X-rays in the neutral
IGM  through relaxation of hydrogen atoms excited by primary
  photoelectrons are boosted by a factor of $f_X$ in each case. For
$f_X=1$ these photons are far less important than Ly-$\alpha$ directly
generated by stars; however, for $f_X\sim \mathcal{O} (100)$ the
contribution of these two sources becomes comparable. In this case,
the 21-cm signal couples to the temperature of the gas
\citep{Wouthuysen:1952, Field:1958} earlier than it would for $f_X
=1$, which can be seen in Figures \ref{fig:2} and \ref{fig:3} for the
cases with soft X-rays and CXRB normalization as well as the extreme
heating case.

One of the striking consequences of the high X-ray efficiency on the
power spectrum is that the amplitude of the 21-cm fluctuations is
suppressed at high redshifts ($\sim15-25$), making their detection
from this epoch harder. In particular, the Ly-$\alpha$ peak at $z\sim
25$ may become unobservable in the case of the soft SED.  This happens
mainly because the Ly-$\alpha$ fluctuations are anti-correlated with
fluctuations seeded by inhomogeneous heating (prior to the heating
transition), which is partially compensated by the enhanced generation
of Ly-$\alpha$ photons by X-ray excitation of the IGM. In most cases,
there remains a significant signal from cosmic dawn observable with
the SKA, even in these worst-case scenarios (in terms of observing
cosmic dawn).  Comparing the three cases of $f_X=1$, $f_X=29.5$
  and $f_X=1000$ with a soft SED, atomic cooling and late EoR, we see
  that the Ly-$\alpha$ peak drops from $95$ mK$^2$ at $z = 22.6$
  ($f_X=1$) to $17$ mK$^2$ at $z = 25.6$ ($f_X= 29.5$) and to $12$
  mK$^2$ at $z = 28.6$ ($f_X=1000$). The effect of X-rays on the
  heating peak in the 21-cm power spectrum is also dramatic. The
  heating peak is lower and is shifted to higher redshifts (lower
  frequencies) when X-rays are strong. In the just-considered case of
  soft X-rays, atomic cooling and late EoR, the heating peak drops
  from $304$ mK$^2$ at $z = 16.9$ ($f_X=1$) to $155$ mK$^2$ at $z =
  21$ ($f_X= 29.5$) to $70$ mK$^2$ at $z = 24.7$ ($f_X=1000$). Thus,
  when the X-ray SED is soft, the normalization to the full unresolved
  CXRB yields a lower limit for the amplitude of the Ly-$\alpha$ and
  X-ray heating peaks in the 21-cm power spectrum and an upper limit
  for the redshift at which these peaks are located. In the case of a
  hard SED, the main effect of X-rays is to suppress the Ly-$\alpha$
  peak, as the fluctuations in temperature are washed out by the large
  mean free paths of the X-ray photons and the heating peak is not
  always present. The intensity of the 21-cm fluctuations from lower
  redshifts ($z\sim 10$) depends on both reionization and heating and,
  thus, is less straightforward to relate to the nature of the X-ray
  sources. In some cases the fluctuations are enhanced, because the
  heating is closer to being saturated during the reionization era; in
  others the fluctuations are suppressed as a result of the X-ray
  contribution to reionization.
 
Heating sources with extremely low X-ray efficiencies do not have
enough time to heat the IGM before the end of reionization, and in
this case the gas remains colder than the CMB even at that time. As a
result, the absorption trough in the global spectrum is very deep, and
the power spectrum features only two peaks sourced by inhomogeneous
Ly-$\alpha$ and ionizing backgrounds. The latter peak is so strong in
the case of $f_X = 0$ that it can be ruled out using current upper
limits on the 21-cm power spectrum at $z = 8.4$ from PAPER as we have
discussed at the end of Section \ref{sec:CXRB}.  The minimal possible
21-cm brightness temperature reaches the value of $T_{21} = -231.7$ mK
($-237.3$ mK) at $z_{\rm min} = 14.2$ ($12.5$) for atomic cooling
(massive halos) with a late EoR and $T_{21} = -218.6$ mK ($-224.6$ mK)
at $z_{\rm min} = 15.2$ (12.8) for atomic cooling (massive halos) with
an early EoR.

 An intriguing result of our study is that, once available, the
  21-cm global signal can be combined with the X-ray background
  produced by high-redshift sources to put constraints on the SED of
  these sources. Because hard sources are more efficient in seeding
  the CXRB, the CXRB limit implies a deeper 21-cm trough (as the
  Universe is colder) than is possible in the case of soft X-ray
  sources. In particular, focusing on our case of the atomic cooling
  halos and late EoR, we see that given the level of the X-ray
  background, the absorption trough in the case of XRBs reaches at
  least $-107.4$ mK, which is much deeper than the corresponding limit
  in the case of the soft X-rays, $-48.8$ mK. Therefore, observing a
  trough shallower than $-107.4$ mK would rule out such a population
  of X-ray sources with a hard SED. In the future, with the direct
  high-redshift X-ray observations of X-ray Surveyor or Athena and the
  observed global 21-cm at hand, it will be possible to significantly
  constrain the properties of the high redshift sources even without
  using any additional information. This method thus provides an
  important alternative to probing the nature of X-ray sources with
  the 21-cm power spectrum \citep{Pritchard:2007, Pacucci:2014,
    Fialkov:2015, Ewall-Wice:2016}. Moreover, having at hand the CXRB
  together with both the global 21-cm signal and its power spectrum
  can help to break degeneracies between X-ray heating and other model
  parameters. We leave a more detailed discussion of this to future
  work.

Finally, based on the whole ensemble of models considered here, it is
possible to define a border line of the parameter space within which
the 21-cm signal can vary without violating any existing observational
constraints. For instance, from the data presented in this paper we
see that the absorption trough of the global signal cannot be
shallower than $-15$ mK or deeper than $-238$ mK, and it is located
between $z = 12.5$ and $z = 24.1$. This border line can be used to
guide future observations. Because the bank of models presented in
this paper is limited,  a more detailed discussion of this
idea can be found in \citet{Cohen:2016} and our future work.

\section{Conclusions}
\label{Sec:sum}

The main results of this paper are the consequences of the upper limit
on the X-ray efficiency which various high-redshift hard or soft X-ray
sources may have without violating the observed soft unresolved X-ray
background.  This limits allows us to estimate the maximal effect that
early X-ray sources can have on the high-redshift 21-cm signal,
reionization, the CMB, and the thermal history of the early Universe.
Our results are particularly useful in light of the existing and
upcoming radio telescopes designed to probe these early epochs.

We find that different types of high-redshift X-ray sources considered
here (i.e., X-ray binaries, hot gas, or mini-quasars) naturally
produce an X-ray background with a slope close to that of the observed
unresolved soft CXRB, which, however, is poorly constrained by present
observations. Therefore, it is hard to reach strong conclusions about
the nature of X-ray sources based on the current measurements of the
shape of the CXRB. On the other hand, normalization to the X-ray
background does constrain the maximum possible X-ray efficiency which
the high redshift heating sources can have. Hard X-ray sources are
more efficient in producing the observed background, compared to the
soft sources, and so the maximal X-ray efficiency is typically around
2--3 times lower for hard sources than for the soft ones. More
generally, the X-ray efficiency that is needed to explain the CXRB is
very high compared to what is normally assumed based on low redshift
observations. This allows for a possibility that the high-redshift
sources may be more luminous than their low-redshift counterparts.
For a population of X-ray sources assumed to have existed down to
redshift 6.2 (and then cut off), this enhancement factor varies
between $\sim 10$ and $\sim 40$, for the various cases; for a
population that only existed at $z=8.5$ and above, the factor is in
the range $\sim 44-190$.

Based on our most extreme case (the early reionization scenario with
massive halos only), we can put a realistic upper bound on the
bolometric luminosity of the high-redshift sources per star formation
rate: $\sim 2.4\times10^{42} ~ \rm{erg\, s}^{-1}\rm{M}_{\odot}^{-1}yr$
for XRBs, $\sim 5.6\times10^{42} ~ \rm{erg\,
  s}^{-1}\rm{M}_{\odot}^{-1}yr$ in the case of hot gas, and $\sim
2.6\times10^{42} ~ \rm{erg\, s}^{-1}\rm{M}_{\odot}^{-1}yr$ for
mini-quasars (assuming the luminosity - SFR relation of
proportionality).  Interestingly, recent observations suggest that
hard X-ray sources such as X-ray binaries or accreting black holes are
a factor of $\sim 10$ more luminous in metal-poor than in
solar-metallicity galaxies \citep{Prestwich:2013, Brorby:2014}. In
addition, X-ray sources in low-redshift dwarf galaxies appear to be
ultra-luminous \citep{Lemons:2015}. However, these trends would have
to greatly increase in order to approach the upper bounds that we
find.

The possible enhanced heating at high redshift has direct implications
for the thermal history and the reionization of the IGM as well as for
the 21-cm signal produced by neutral hydrogen during and prior to
cosmic reionization. In particular, cosmic heating can happen up to
$\Delta z \sim 4-5$ earlier than what is normally assumed. 
  However, even with the enhanced X-ray emission the saturated heating
  assumption throughout reionization can only be used with soft
  sources. In addition, efficient X-ray sources compete with stellar
sources in ionizing the gas. In particular, X-ray sources with soft
spectra can be so efficient as to compete with UV photons in
reionizing the Universe, producing a fractional increase in the
optical depth to CMB photons of $12- 16 \%$. In the cases when X-rays
are so efficient, reionization happens more homogeneously since X-rays
have a very long mean free path compared to UV and, thus, ionize the
IGM far from the sources. On the other hand, hard sources such as XRBs
and mini-quasars (with their maximal normalization) have only a mild
effect on reionization, producing a fractional increase in the CMB
optical depth of only $1-3 \%$.

Thus, in our model reionization is mainly driven by stellar sources
which emit UV, while quasars can have a small impact on the EoR
channeling their energy into hard X-rays. A different situation was
described recently by \citet{Madau:2015} who explored a possibility of
reionization achieved only by high redshift quasars and other faint,
high-redshift AGN. \citet{Madau:2015} assigned a composite emissivity
to their quasars accounting for a broad spectrum of emitted photons,
from optical to hard X-rays \citep{Haardt:2012}, which resulted in a
stronger contribution of AGN to reionization. These authors found that
$ z > 5$ active galaxies can reionize the universe without
overproducing the unresolved CXRB at 2 keV (amounting to $\sim 60 \%$
of the CXRB) provided their properties are similar to those of their
lower redshift counterparts. The contribution of high redshift quasars
in this case was normalized by fitting the observed integrated optical
emissivity at redshifts up to $\sim 5$ \citep{Bongiorno:2007,
  Schulze:2009, Giallongo:2015} and extrapolating this result to
higher redshifts out to $z\sim 12$. This emissivity drops slowly with
redshift, predicting a significant population of quasars at early
times. Based as it is on the extrapolation of current scarce
high-redshift observations, this result may overestimate the role of
quasars and AGN in reionization.

An important implication of the possible high X-ray efficiency of the
early X-ray sources is the signature that they imprint in the 21-cm
signal of neutral hydrogen. This signal has not been detected yet, and
its discovery is a major goal of the astronomical community. Since the
agents of high-redshift cosmic heating are highly unconstrained, a
wide space of possibilities is left for the 21-cm signal, which
translates into a wide parameter space for telescopes to search.
Having an upper limit on the effects of cosmic heating naturally
constrains the search parameter space and can help guide telescope
design and search strategies. An upper limit on heating has important
implications both for the global spectrum and for the spatial
fluctuations of the 21-cm signal.

In the case of a high X-ray efficiency, the absorption trough of the
global 21-cm signal is significantly reduced and shifted towards lower
frequencies (higher redshifts), while the reionization gradient (in
emission) is extended over a broader range of frequencies. The 21-cm
power spectrum is suppressed at high redshifts ($z\sim 15-25$) due to
the enhanced heating, and the Ly-$\alpha$ peak is reduced and shifted
to lower frequencies as well as the X-ray peak (if present); however,
the fluctuations from the second half of reionization are maintained
and sometimes boosted due to the effect of saturated heating. Thus,
there is mixed news for 21-cm experiments.  For global
  experiments, the expected signal from reionization and cosmic dawn
  is bounded and for some models can be significantly reduced. For
phase 1 of the SKA, reionization remains observable at a high
signal-to-noise ratio in all cases, but cosmic dawn can become
difficult to observe in the most extreme cases. We note that in our
calculation we do not account for exotic processes, such as dark
matter annihilation, which could heat up the gas early on and suppress
the 21-cm signal without leaving any X-ray background.

For completeness, we have also considered lower limits on the X-ray
heating efficiency for late reionization scenarios (which do not end
before $z=8.4$) using recent upper limit found by PAPER. Weak or no
X-ray heating results in a 21-cm signal with a deep absorption trough
in the global signal and strong fluctuations from all the epochs that
normally  have a signal that is suppressed due to cosmic heating. Such a signal should be an
easy target for radio telescopes. 

 To summarize, the unknown details of star formation, the process
  of reionization and the nature of high-redshift X-ray sources,
  including their SED and X-ray efficiency, result in a large
  uncertainty in the expected 21-cm signal.  Specifically, the
  absorption trough, the main feature of the global 21-cm signal, can
  vary in depth anywhere from $T_{21}\sim -240$~mK to $-15$~mK, and in
  position from $z \sim 12$ to 24; the fluctuation peak from
  mid-reionization is still unconstrained to better than the range of
  3-3000~mK$^2$ at $k = 0.2$ Mpc$^{-1}$; the X-ray peak strongly
  depends on the properties of the X-ray sources and can be strong,
  mild or vanishing; finally, the strength of the fluctuation peak
  from the Ly-$\alpha$ coupling era varies between 2 and 200 ~mK$^2$
  and its redshift lies within $z \sim 13-25$.

While the CXRB narrows the possible realization space of the 21-cm
background only slightly at present, future X-ray missions such as
Athena and X-ray Surveyor could directly probe high redshift sources
of X-rays, constraining their effect on the intergalactic gas and the
21-cm signal.

\section{Acknowledgments}
We thank L. V. E. Koopmans for providing the noise spectrum of SKA and
A, Vikhlinin for his input on the CXRB.  We also thank the
  anonymous referee for carefully reading our work and providing
  valuable comments.

A. F.  was partially supported by the LabEx ENS-ICFP: ANR-10-
LABX-0010/ANR-10-IDEX- 0001-02 PSL.  R.B.\ and A.C.\ acknowledge
Israel Science Foundation grant 823/09 and the Ministry of Science and
Technology, Israel. R.B.'s work has been done within the Labex
Institut Lagrange de Paris (ILP, reference ANR-10-LABX-63) part of the
Idex SUPER, and received financial state aid managed by the Agence
Nationale de la Recherche, as part of the programme Investissements
d'avenir under the reference ANR-11-IDEX-0004-02. R.B. also
acknowledges a Leverhulme Trust Visiting Professorship; this research
was also supported in part by Perimeter Institute for Theoretical
Physics. Research at Perimeter Institute is supported by the
Government of Canada through Industry Canada and by the Province of
Ontario through the Ministry of Economic Development \& Innovation.
J.S.\ was supported by the ERC Project No. 267117 (DARK) hosted by
Universit´e Pierre et Marie Curie (UPMC) - Paris 6, PI J. Silk. JS
acknowledges the support of the JHU by NSF grant OIA-1124403.



\begin{thebibliography}{14}
\bibitem[\protect\citeauthoryear{Ade et al.}{2013}]{Planck:2013}
Ade, P. A. R., et al., 2014, A\&A, 571, 16


\bibitem[\protect\citeauthoryear{Ade et al.}{2015}]{Planck:2015}
Ade, P. A. R., et al., 2016, A\&A, 594, 13

\bibitem[\protect\citeauthoryear{Adam et al.}{2016}]{Planck:2016}
Adam, R., et al., 2016, arXiv:1605.03507

 \bibitem[\protect\citeauthoryear{Ali et al.}{2015}]{Ali:2015}
	Ali, Z. S., Parsons, A. R., Zheng, H., Pober, J. C., Liu, A., et al. 2015, ApJ, 809, 61
	
	\bibitem[\protect\citeauthoryear{Barkana \& Loeb}{2004}]{Barkana:2004}
Barkana R., Loeb A., 2004, ApJ, 609, 474

\bibitem[\protect\citeauthoryear{Barkana \& Loeb}{2005}]{Barkana:2005}
Barkana, R., \& Loeb, A. 2005, ApJ, 626, 1

\bibitem[\protect\citeauthoryear{Bauer et al.}{2004}]{Bauer:2004}
Bauer, F. E., Alexander, D. M., Brandt, W. N., et al., 2004, AJ, 128, 2048


\bibitem[\protect\citeauthoryear{Becker et al.}{2015}]{Becker:2015} 
Becker, G. D., Bolton, J. S., Madau, P., et al., 2015, MNRAS, 447, 3402

\bibitem[\protect\citeauthoryear{Bernardi et al.}{2016}]{Bernardi:2016} 
Bernardi, G., Zwart, J.~T.~L., Price, D., et al.\ 2016, MNRAS, 461, 2847 

\bibitem[\protect\citeauthoryear{Bongiorno et al.}{2007}]{Bongiorno:2007}
Bongiorno, A., Zamorani, G., Gavignaud, I., et al., 2007, A\&A
472, 443

\bibitem[\protect\citeauthoryear{Bowman \& Rogers}{2010}]{Bowman:2010}
Bowman, J. D., \& Rogers, A. E. E., 2010, Nature, 468, 796 

\bibitem[\protect\citeauthoryear{Bowman et al.}{2013}]{Bowman:2013}
Bowman, J. D., Cairns, I., Kaplan D. L., Murphy, T., Oberoi, D.,  et al., 2013, PASA, 30, 031


\bibitem[\protect\citeauthoryear{Brorby et al.}{2014}]{Brorby:2014}
Brorby, M., Kaaret, P., \& Prestwich, A. 2014, MNRAS, 441, 2346

\bibitem[\protect\citeauthoryear{Burns et al.}{2012}]{Burns:2012}
Burns, J. O., Lazio, J., Bale, S., Bowman, J., Bradley, R., et al., 2012, AdSpR, 49, 433

\bibitem[\protect\citeauthoryear{Cappelluti et al.}{2012}]{Cappelluti:2012}
Cappelluti, N., Ranalli, P., Roncarelli, M., Arevalo, P., Zamorani, G., et al., 2012, MNRAS, 427, 651


\bibitem[\protect\citeauthoryear{Christian \& Loeb}{2013}]{Christian:2013}
Christian, P., Loeb, A., 2013, JCAP, 09, 014

\bibitem[\protect\citeauthoryear{Cirelli et al.}{2009}]{Cirelli:2009}
Cirelli, M., Iocco, F., Panci, P., 2009, JCAP, 10, 009

 
\bibitem[\protect\citeauthoryear{Cohen et al.}{2016a}]{Cohen:2015}
Cohen, A., Fialkov, A., Barkana, R., 2016a, MNRAS, 459, 90
 
 \bibitem[\protect\citeauthoryear{Cohen et al.}{2016b}]{Cohen:2016}
Cohen, A., Fialkov, A., Barkana, R., 2016b, arXiv:1609.02312


\bibitem[\protect\citeauthoryear{Dijkstra et al.}{2004}]{Dijkstra:2004}
Dijkstra, M., Haiman, Z., \& Loeb, A. 2004, MNRAS, 613, 646

\bibitem[\protect\citeauthoryear{Dijkstra et al.}{2012}]{Dijkstra:2012}
Dijkstra, M., Gilfanov, M., Loeb, A., Sunyaev, R., 2012, MNRAS, 421, 213

\bibitem[\protect\citeauthoryear{Ewall-Wice et al.}{2016}]{Ewall-Wice:2016}
Ewall-Wice, A., Hewitt, J., Mesinger, A., Dillon, J. S., Liu, A., Pober, J., 2016, MNRAS, 458, 2710

\bibitem[\protect\citeauthoryear{Fialkov et al.}{2014}]{Fialkov:2014}
Fialkov, A., Barkana, R., Visbal, E., 2014, Nature, 506, 197

\bibitem[\protect\citeauthoryear{Fialkov \& Barkana}{2014}]{Fialkov:2014b}
Fialkov, A. \& Barkana, R.,  2014, MNRAS, 445, 213

\bibitem[\protect\citeauthoryear{Fialkov}{2014}]{Fialkov:2014c}
Fialkov, A., 2014, IJMPD, 2330017



\bibitem[\protect\citeauthoryear{Fialkov et al.}{2015}]{Fialkov:2015}
Fialkov, A., Barkana, R. \& Cohen, A., 2015, PRL, 114, 1303


 \bibitem[\protect\citeauthoryear{Field}{1958}]{Field:1958}
 Field, G. B., 1958, Proc. Institute Radio Energeers, 46, 240
 
 
\bibitem[\protect\citeauthoryear{Fragos et al.}{2013}]{Fragos:2013}
Fragos, T., Lehmer, B. D., Naoz, S., Zezas, A., Basu-Zych, A., 2013, ApJ, 776, 31

\bibitem[\protect\citeauthoryear{Furlanetto et al.}{2004}]{Furlanetto:2004}
Furlanetto, S. R., Zaldarriaga, M.,  Hernquist, L., 2004, ApJ, 613, 1


\bibitem[\protect\citeauthoryear{Furlanetto et al.}{2006}]{Furlanetto:2006}
Furlanetto, S. R., Oh, S. P., Briggs, F. H., 2006, PhR, 433, 181

\bibitem[\protect\citeauthoryear{Furlanetto}{2006}]{Furlanetto:2006b}
Furlanetto, S. R., 2006, MNRAS 371, 867


\bibitem[\protect\citeauthoryear{Furlanetto \& Stoever}{2010}]{Furlanetto:2010}
Furlanetto, S. R., Stoever, S. J.,  2010, MNRAS 404, 1869


\bibitem[\protect\citeauthoryear{Giacconi et al.}{1962}]{Giacconi:1962}
Giacconi, R., Gursky, H., Paolini, F., Rossi, B., 1962, PRL, 9, 439

\bibitem[\protect\citeauthoryear{Giallongo et al.}{2015}]{Giallongo:2015}
Giallongo, E., Grazian, A., Fiore, F., et al., 2015, A\&A 578, A83

\bibitem[\protect\citeauthoryear{Grimm et al.}{2003}]{Grimm:2003}
	Grimm, H.-J., Gilfanov, M., Sunyaev, R., 2003, MNRAS, 339, 793
	
	\bibitem[\protect\citeauthoryear{Gilfanov et al.}{2004}]{Gilfanov:2004}
	Gilfanov, M., Grimm, H.-J., Sunyaev, R.,  2004, MNRAS, 347, 57 

\bibitem[\protect\citeauthoryear{Haardt \& Madau}{2012}]{Haardt:2012}
Haardt, F. \& Madau, P., 2012, ApJ, 746, 125

\bibitem[\protect\citeauthoryear{Hasinger et al.}{1993}]{Hasinger:1993}
Hasinger, G., Burg, R., Giacconi, R., Hartner, G., Schmidt, et al., 1993, A\&A, 275, 1

\bibitem[\protect\citeauthoryear{Hickox \& Markevich}{2006}]{Hickox:2006}
Hickox R. C., Markevitch M., 2006, ApJ, 645, 95

\bibitem[\protect\citeauthoryear{Koopmans et al.}{2015}]{Koopmans:2015}
Koopmans, L., Pritchard, J., Mellema, G., Aguirre, J., Ahn, K., et al., 2015, AASKA14, 1

\bibitem[\protect\citeauthoryear{Liu et al.}{2013}]{Liu:2013}
Liu, A., Pritchard, J. R., Tegmark, M., Loeb, A., 2013, PRD, 87, 3002

\bibitem[\protect\citeauthoryear{Leitherer et al.}{1999}]{Leitherer:1999}
Leitherer, C., Schaerer, D., Goldader, J. D., Delgado, R. M. G., Robert, C., et al., 1999, ApJS, 123, 3

\bibitem[\protect\citeauthoryear{Lehmer et al.}{2012}]{Lehmer:2012}
Lehmer, B. D., Xue, Y. Q., Brandt, W. N., Alexander, D. M., Bauer, F. E., et al., 2012, ApJ, 752, 46L

\bibitem[\protect\citeauthoryear{Lemons et al.}{2015}]{Lemons:2015}
Lemons, S. M., Reines, A. E., Plotkin, R. M., Gallo, E., Greene, J. E., 2015, ApJ, 805, 12


\bibitem[\protect\citeauthoryear{Madau et al.}{2004}]{Madau:2004}
Madau, P., Rees, M. J., Volonteri, M., Haardt, F., Oh, S. P., 2004, ApJ, 604, 484

\bibitem[\protect\citeauthoryear{Madau \& Haardt}{2015}]{Madau:2015}
Madau, P. \& Haardt, F., 2015, ApJ, 813, 8


\bibitem[\protect\citeauthoryear{Majumdar et al.}{2015}]{Majumdar:2015}
Majumdar, S., Jensen, H., Mellema, G., Chapman, E., Abdala F. B., 2016, MNRAS, 456, 2080


\bibitem[\protect\citeauthoryear{Marleau et al.}{2014}]{Marleau:2014}
Marleau, F. R., Clancy, D., Bianconi, M., Habas, R., 2014, arXiv:1411.3844


\bibitem[\protect\citeauthoryear{McCammon et al.}{2002}]{McCammon:2002}
McCammon, D., Almy, R., Apodaca, E., Bergmann Tiest, W., Cui, W.,  et al., 2002, ApJ, 576, 188

\bibitem[\protect\citeauthoryear{McQuinn}{2012}]{McQuinn:2012}
McQuinn, M., 2012, MNRAS, 426, 1349

\bibitem[\protect\citeauthoryear{Mesinger et al.}{2013}]{Mesinger:2013}
Mesinger, A., Ferrara, A., Spiegel, D. S., 2013, MNRAS, 431, 621

\bibitem[\protect\citeauthoryear{Mineo et al.}{2012a}]{Mineo:2012a}
Mineo, S., Gilfanov, M., Sunyaev, R., 2012a, MNRAS, 426, 1870 

\bibitem[\protect\citeauthoryear{Mineo et al.}{2012b}]{Mineo:2012b}
Mineo, S., Gilfanov, M., Sunyaev, R., 2012b, MNRAS, 419, 2095 

\bibitem[\protect\citeauthoryear{Mirabel et al.}{2011}]{Mirabel:2011}
Mirabel, I. F., Dijkstra, M., Laurent, P., Loeb, A. \& Pritchard, J. R., 2011, A\& A 528, A149

\bibitem[\protect\citeauthoryear{Mirocha}{2014}]{Mirocha:2014}
Mirocha, J.,  2014, MNRAS, 443, 1211

\bibitem[\protect\citeauthoryear{Moran et al.}{2014}]{Moran:2014}
Moran, E. C., Shahinyan, K., Sugarman, H. R., Velez, D. O., Eracleous, M., 2014, AJ, 148, 136

\bibitem[\protect\citeauthoryear{Morandi \& Barkana}{2012}]{Morandi}
Morandi, A., Barkana, R., 2012, MNRAS, 424, 2551

\bibitem[\protect\citeauthoryear{Moretti}{2012}]{moretti} Moretti, A.,
  Vattakunnel, S., Tozzi, P., et al.\ 2012, A\&A, 548, A87

\bibitem[\protect\citeauthoryear{Oh}{2001}]{Oh:2001}
Oh, S. P., 2001, ApJ, 553, 499


\bibitem[\protect\citeauthoryear{Paciga et al.}{2013}]{Paciga:2013}
Paciga, G., Albert, J. G., Bandura, K., Chang, T.-C., Gupta, Y., et al., 2013, MNRAS, 433, 639


\bibitem[\protect\citeauthoryear{Pacucci et al.}{2014}]{Pacucci:2014}
Pacucci, F., Mesinger, A., Mineo, S., Ferrara, A.,  et al., 2014, MNRAS, 443, 678


\bibitem[\protect\citeauthoryear{Parsons et al.}{2014}]{Parsons:2014}
Parsons, A. R., Liu, A., Aguirre, J. E., Ali, Z. S., Bradley, R. F., et al., 2014, ApJ, 788, 106  


   
      \bibitem[Patra et al. (2013)]{Patra:2013}  Patra, N., Subrahmanyan, R., Raghunathan, A., Udaya Shankar, N., 2013, ExA, 36, 319
   
   
  \bibitem[Pober et al. (2015)]{Pober:2015}   Pober, J. C., Ali, Z. S., Parsons, A. R., 
   McQuinn, M., Aguirre, J. E., et al.  2015, ApJ 809, 62
   
    \bibitem[Power et al. (2009)]{Power:2009}  Power, C., Wynn, G. A., Combet, C., Wilkinson, M. I., 2009, MNRAS, 395, 1146
   
    \bibitem[Power et al. (2013)]{Power:2013}  Power, C., James, G., Combet, C., Wynn, G.,  2013, ApJ, 764, 76 

\bibitem[\protect\citeauthoryear{Prestwich et al.}{2013}]{Prestwich:2013}
Prestwich, A. H., Tsantaki, M., Zezas, A., et al., 2013, ApJ, 769, 92

\bibitem[\protect\citeauthoryear{Pritchard \& Furlanetto}{2007}]{Pritchard:2007}
Pritchard, J. R. \& Furlanetto, S. R. 2007, MNRAS, 376, 1680

\bibitem[\protect\citeauthoryear{Pritchard \& Loeb}{2008}]{Pritchard:2008}
Pritchard, J. R. \& Loeb, A., 2008, PRD, 78, 3511

\bibitem[\protect\citeauthoryear{Pritchard \& Loeb}{2010}]{Pritchard:2010}
Pritchard, J. R. \& Loeb, A., 2010, PRD, 82, 023006

\bibitem[\protect\citeauthoryear{Pritchard \& Loeb}{2012}]{Pritchard:2012}
Pritchard, J. R. \& Loeb, A., 2012, RPP, 75, 6901

\bibitem[\protect\citeauthoryear{Ripamonti et al.}{2008}]{Ripamonti:2008}
	Ripamonti, E., Mapelli, M., Zaroubi, S., 2008, MNRAS, 387, 158
	
	
\bibitem[\protect\citeauthoryear{Schulze}{2009}]{Schulze:2009}
Schulze, A., Wisotzki, L., \& Husemann, B., 2009, A\&A 507, 781

\bibitem[\protect\citeauthoryear{Shakura \&
    Sunyaev}{1973}]{Shakura:1973} Shakura N. I., Sunyaev R. A., 1973,
  A\&A, 24, 337

\bibitem[\protect\citeauthoryear{Shaver}{1999}]{Shaver} Shaver,
  P.\ A., Windhorst, R.\ A., Madau, P., de Bruyn, A.\ G., 1999, A\&A,
  345, 380

\bibitem[\protect\citeauthoryear{Tanaka et al.}{2012}]{Tanaka:2012}
Tanaka, T., Perna, R., Haiman, Z.,2012, MNRAS, 425, 2974


\bibitem[\protect\citeauthoryear{Tseliakhovich et al.}{2011}]{Tseliakhovich:2011}
Tseliakhovich D., Barkana R., Hirata C. M. 2011, MNRAS,
418, 906


\bibitem[\protect\citeauthoryear{Tseliakhovich \& Hirata}{2010}]{Tseliakhovich:2010}
Tseliakhovich D., Hirata C. M., 2010, PRD, 82, 083520

\bibitem[\protect\citeauthoryear{van Haarlem  et al.}{2013}]{Haarlem:2013}
van Haarlem, M. P., Wise, M. W., Gunst, A. W., Heald, G., et al., 2013, A\&A 556, 2

\bibitem[\protect\citeauthoryear{Visbal et al.}{2012}]{Visbal:2012}
Visbal, E., Barkana, R., Fialkov, A., Tseliakhovich, D., \&
Hirata, C. M., 2012, Nature, 487, 70

\bibitem[\protect\citeauthoryear{Voytek et al.}{2014}]{Voytek:2014}
  Voytek, T.~C., Natarajan, A., J{\'a}uregui Garc{\'{\i}}a, J.~M.,
  Peterson, J.~B., \& L{\'o}pez-Cruz, O.\ 2014, ApJL, 782, L9

\bibitem[\protect\citeauthoryear{Weisskopf et al.}{2015}]{Weisskopf:2015}
Weisskopf, M. C., Gaskin, J., Tananbaum, H. \& Vikhlinin, A., 2015, SPIE, 9510, 02

\bibitem[\protect\citeauthoryear{Wouthuysen}{1952}]{Wouthuysen:1952}
Wouthuysen, S. A., 1952, AJ, 57, 31


\bibitem[\protect\citeauthoryear{Wyithe \& Loeb}{2003}]{Wyithe:2003}
Wyithe, J. S. B., Loeb, A., 2003, ApJ, 595, 614

\bibitem[\protect\citeauthoryear{Wyithe \& Loeb}{2003}]{Wyithe:2003b}
Wyithe, J. S. B., Loeb, A., 2003, ApJ, 586, 693

\bibitem[\protect\citeauthoryear{Xue et al.}{2011}]{Xue:2011}
Xue, Y. Q., Luo, B., Brandt, W. N., et al., 2011, ApJS, 195, 10

\bibitem[Zarka et al. (2012)]{NenuFAR} 
Zarka, P., Girard, J. N., Tagger, M., \& Denis, L. 2012, SF2A-2012: Proceedings of the Annual meeting of the French Society of Astronomy and Astrophysics

\label{lastpage}

\end{thebibliography}
\end{document}